\documentclass[aps, prd, onecolumn, tightenlines, notitlepage, superscriptaddress, nofootinbib, preprintnumbers, floatfix,showkeys,11pt,altaffilletter]{revtex4-2}

\usepackage[normalem]{ulem}
\usepackage{amstext}
\usepackage{amssymb}
\usepackage{amsmath}
\usepackage{bm}
\usepackage{graphicx}
\graphicspath{{plots/}}
\usepackage{url}
\usepackage{color}
\usepackage{ulem}
\usepackage[utf8]{inputenc}
\usepackage{ragged2e}
\pdfoutput=1
\usepackage{textcomp}
\usepackage{comment}
\usepackage{wasysym}   \newcommand{\WWcross}{{\raisebox{-1pt}{\rotatebox{90}{\Bowtie}}}}
\usepackage{yfonts}
\usepackage{epsfig,amsfonts,mathrsfs,amsmath,amssymb,graphicx,color,slashed,multirow}
\usepackage{amsmath,latexsym,amssymb,graphicx,slashed,color,enumerate,url,cancel,gensymb}
\usepackage{textcomp}

\usepackage[x11names]{xcolor}
\usepackage[colorlinks]{hyperref}

\usepackage{textcase}
\usepackage{amsmath}
\usepackage{booktabs}
\usepackage{adjustbox}

\definecolor{vdrgreen}{rgb}{0.0, 0.7, 0.0}

\definecolor{indianred}{rgb}{0.8, 0.36, 0.36}
\definecolor{blue(ncs)}{rgb}{0.0, 0.53, 0.74}
\AtBeginDocument{\hypersetup{citecolor=indianred,linkcolor=indianred,urlcolor=indianred}}

\usepackage{float}

\usepackage{lmodern}
\usepackage{ae,aecompl}
\usepackage{appendix}
\usepackage{orcidlink}
\usepackage{nccmath} % Provides command \medmath
\usepackage{textcomp}
\usepackage{upgreek}
\usepackage[version=4]{mhchem}
\usepackage[utf8]{inputenc}
\usepackage{fontawesome}
\usepackage{yfonts}
\usepackage[official]{eurosym}
\usepackage{textgreek}


\newcommand{\diff}{\mathrm{d}}
\def\cevns{CE\textnu NS~}
\def\eves{E\textnu ES~}

\newcommand{\qtransfer}{\left|\mathbf{q}\right|}

\usepackage[T1]{fontenc}
\usepackage{ae,aecompl}
\graphicspath{{Figures/}}
\usepackage{appendix}
\usepackage{placeins}
\usepackage{subfigure}



\newcommand{\AddrIISERB}{Department of Physics, Indian Institute of Science Education and Research - Bhopal, \\ 
Bhopal Bypass Road, Bhauri, Bhopal 462066, India}

\newcommand{\AddrIITBHU}{%
Department of Physics, Indian Institute of Technology (BHU), Varanasi 221005, India}

\usepackage{orcidlink}
\begin{document}

\title{\Large Precision tests of SM and new physics with the COHERENT Ge-mini \\ and TEXONO data}

\author{Tousib Ahmed~\orcidlink{0009-0004-1830-5619}}\email{tousibahmed.phy24@itbhu.ac.in}
\affiliation{\AddrIITBHU}
\author{Ayan Chattaraj~\orcidlink{0009-0001-3561-7049}}
\email{ayan23@iiserb.ac.in}
\affiliation{\AddrIISERB}
\author{Anirban Majumdar~\orcidlink{0000-0002-1229-7951}}\email{anirban19@iiserb.ac.in}
\affiliation{\AddrIISERB}
\author{Newton Nath~\orcidlink{0000-0002-0592-0020}}\email{nnath.phy@iitbhu.ac.in}
\affiliation{\AddrIITBHU}
\author{Rahul Srivastava~\orcidlink{0000-0001-7023-5727}}\email{rahul@iiserb.ac.in}
\affiliation{\AddrIISERB}


%%%%%%%%%%%%%%%%%%%%%%%%%%%%%%%%%%%%%%%%%%%%%%%%%%%%%
\begin{abstract}
A comprehensive numerical analysis of the latest germanium based coherent elastic neutrino nucleus scattering (CE\textnu NS) data from the COHERENT Ge-mini and TEXONO experiments has been conducted to test the Standard Model (SM) and search for new physics.
By combining CE\textnu NS and E\textnu ES signals with a consistent treatment of detector effects and systematic uncertainties, we obtain a low energy determination of the weak mixing angle from COHERENT Ge-mini, in agreement with the SM prediction.
We derive novel constraints on neutrino electromagnetic properties, including the magnetic moment, millicharge, charge radius, and anapole moment, with TEXONO providing particularly competitive bounds.
Inclusion of E\textnu ES significantly improves the sensitivity to the neutrino millicharge by up to three orders of magnitude.
We also investigated light scalar and vector mediators, finding striking complementarity between reactor and stopped pion sources across different mediator mass regimes.
Finally, we put bounds on sterile neutral lepton production through transition dipole, scalar, and vector portals, probing masses from the sub-MeV to tens of MeV scale.
Our results demonstrate that current germanium based CE\textnu NS experiments provide a powerful low energy laboratory for precision electroweak tests and complementary probes of a broad class of physics beyond the SM.
\end{abstract}

\maketitle
\section{Introduction}
\label{sec:intro}
The Standard Model (SM) of particle physics provides a remarkably successful 
description of the known elementary particles and their interactions. 
Nevertheless, several observations point toward deviations that call for an 
extension of this framework. One of the most compelling pieces of evidence of 
physics beyond the Standard Model (BSM) is the neutrino mass, a discovery that 
led to the 2015 Nobel Prize in Physics~\cite{Super-Kamiokande:1998kpq, SNO:2001kpb}.
However, several other neutrino properties, such as their electromagnetic interactions 
including the charge radius, millicharge, and magnetic moment, neutrino 
generalized interactions, and sterile neutrino upscattering, are also considered 
promising signatures of BSM physics,
either through the observation 
of SM forbidden processes or through anomalously large values of otherwise 
SM allowed observables.
Owing to rapid advances in neutrino detection technology, current and next generation neutrino experiments are placing increasing emphasis on probing these elusive neutrino properties with unprecedented precision.
Among the various neutrino probes, experiments that seek Coherent Elastic Neutrino Nucleus Scattering (CE\textnu NS) have attracted considerable attention as a powerful tool to probe both SM and BSM physics.
In this process, a neutrino scatters coherently off the entire nucleus, inducing a nuclear ground state to ground state transition, provided that the momentum transfer remains comparable to or smaller than the inverse of the nuclear radius. Under this condition, the contributions of all nucleons add up coherently, resulting in an enhanced cross section that scales approximately with the square of the neutron number of the target nucleus. Owing to this coherent enhancement, \cevns attains the largest cross section among the relevant SM neutrino interactions at low energies, significantly surpassing other relevant processes such as elastic neutrino electron scattering (E\textnu ES).

CE\textnu NS, originally proposed by Freedman in 1974~\cite{Freedman:1973yd}, 
with a later independent proposal by Drukier and Stodolsky in 
1984~\cite{Drukier:1984vhf}, was first observed nearly four decades later, in 
2017, by the COHERENT collaboration using a CsI detector at the Oak Ridge 
National Laboratory (ORNL), exploiting neutrinos from the Spallation Neutron 
Source (SNS)~\cite{COHERENT:2017ipa}.
Experimental detection of such processes was a mammoth task due to the small nuclear recoil energies involved.
Subsequently, such neutrino events have been firmly established by the same 
collaboration using other detector targets, such as liquid argon 
(LAr)~\cite{COHERENT:2020iec} in 2020 and recently using germanium (Ge)~\cite{COHERENT:2024axu} in 2024.
Meanwhile, reactor based experiments such as Dresden-II~\cite{Colaresi:2022obx} in 2022 and CONUS+~\cite{Ackermann:2025obx} in 2025 have also announced their CE\textnu NS observation using Ge detectors. Moreover, \cevns signals induced by the solar $^8$B neutrino flux have been reported by dark matter direct detection experiments such as XENONnT, PandaX-4T, and LUX-ZEPLIN using liquid Xe detectors~\cite{PandaX:2024muv,XENON:2024ijk,LZ:2025igz,XENON:2026ydt}.
For a latest review see Ref.~\cite{Abdullah:2022zue}.
Since the first detection, 
\cevns  datasets have been widely exploited to search for a variety of new physics scenarios, like constraining the electromagnetic properties of neutrinos~\cite{Cadeddu:2019eta, AtzoriCorona:2022qrf, DeRomeri:2022twg, A:2022acy, DeRomeri:2024hvc, Chattaraj:2025fvx, DeRomeri:2025csu, AtzoriCorona:2026wbu}, 
neutrino generalized interactions~\cite{Dutta:2015nlo,Lindner:2016wff,AristizabalSierra:2018eqm,Flores:2021kzl, Majumdar:2021vdw, Majumdar:2022nby,  Chattaraj:2025rtj}, new interactions mediated by light~\cite{Farzan:2018gtr,Denton:2018xmq,Flores:2020lji,Cadeddu:2020nbr, delaVega:2021wpx, Bertuzzo:2021opb, AristizabalSierra:2022axl, Majumdar:2024dms,Xia:2024ytb, DeRomeri:2024iaw, Blanco-Mas:2024ale, Chattaraj:2025fvx, DeRomeri:2025csu, DeRomeri:2026prc} as well as heavy~\cite{Billard:2018jnl,Arcadi:2019uif, DeRomeri:2026prc} scalar and vector mediators, sterile neutrino oscillations and deviations from unitarity of the leptonic mixing 
matrix~\cite{Majumdar:2022nby, CentellesChulia:2024sff, Chattaraj:2025rtj, CentellesChulia:2025jir}, the upscattering production of sterile neutral leptons~\cite{Miranda:2021kre,  Candela:2023rvt, Candela:2024ljb, DeRomeri:2024hvc, Chattaraj:2025rtj, DeRomeri:2025csu}, among others.

Recently, the COHERENT collaboration achieved the most precise measurement of the CE\textnu NS cross section to date with the germanium detector array, Ge-mini, at the ORNL~\cite{COHERENT:2026yje}.
To measure nuclear recoils, they used  four inverted coaxial point contact high purity germanium (ICPC HPGe) detectors, named Ge-21, Ge-25, Ge-26, and Ge-28, operated in SNS with sub-keV energy thresholds~\cite{COHERENT:2026yje,Bouabid:2025upo}.  So far, the collaboration has published datasets for two periods. The first measurement was made between June and August 2023~\cite{COHERENT:2024axu}, and the second measurement was made between February and May 2025~\cite{COHERENT:2026yje}.
In addition, the TEXONO experiment, which is capable of measuring the low energy and intense flux of electron antineutrinos from nuclear reactors, reported their latest results using  germanium detectors at the Kuo-Sheng Reactor Neutrino Laboratory (KSNL) \cite{TEXONO:2026eyr}.
To find reactor based CE\textnu NS, the collaboration used two electro-cooled point-contact germanium (ECPCGe) detectors, denoted D70 and D50, in KSNL \cite{TEXONO:2026eyr}.
Using the latest COHERENT Ge-mini data, Ref.~\cite{AtzoriCorona:2026wbu} determined the weak mixing angle, the neutrino charge radii, and the neutron root mean square radius of germanium nuclei, and derived bounds on nonstandard neutrino interactions.
Subsequently, Ref.~\cite{DeRomeri:2026dac} performed an analysis of this dataset to determine the germanium neutron radius and neutron skin, as well as the low energy weak mixing angle.

In this work, we performed an extensive numerical analysis of the latest CE\textnu NS datasets published by  the COHERENT ~\cite{COHERENT:2026yje} and TEXONO ~\cite{TEXONO:2026eyr} collaborations to investigate both the SM and  a broad class of BSM physics. For the SM, we put constraint on the weak mixing angle, $\sin^2\theta_W$, one of the fundamental parameters of electroweak theory. 
The primary focus of this work is to investigate  various BSM scenarios by combining both \cevns and E\textnu ES channels of COHERENT Ge-mini and TEXONO. 
The BSM framework includes the electromagnetic properties of neutrinos like the effective neutrino magnetic moment, millicharge, charge radius, and anapole moment, which provide sensitive probes of new physics beyond the SM.
Moreover, we also examined BSM scenarios such as neutrino generalized interactions mediated by light scalar and vector bosons, with the vector interaction realized through the anomaly free $U(1)_{B\!-\!L}$ gauge extension.
 In addition, sterile neutrino upscattering is explored through both the transition dipole portal and generalized scalar and vector interactions, allowing active neutrinos to scatter into heavier sterile neutral leptons. 
 By incorporating both nuclear and electron recoil signals in a unified analysis, this work significantly enhances the sensitivity to these SM and BSM physics and derives competitive constraints on neutrino electromagnetic properties, light mediators, neutrino generalized interactions (NGIs) and sterile neutrino properties. 

 The paper is organized as follows: in Sec.~\ref{Sec:Theory}, we present the theoretical framework, outlining the \cevns and \eves differential cross sections within the SM as well as for the various BSM scenarios. Sec.~\ref{Sec:Events} describes the experimental configurations of the COHERENT Ge-mini and TEXONO detectors, the corresponding event rate simulations, and the statistical procedure adopted in our analysis. In Sec.~\ref{Sec:Results}, we present and discuss our results, deriving constraints on the weak mixing angle and on the different BSM scenarios, and compare them with existing limits from other terrestrial, astrophysical, and cosmological probes. Finally, we summarize our main findings and conclude in Sec.~\ref{Sec:Conclusions}.

\section{\label{Sec:Theory} Theoretical framework}
In this section, we outline the \cevns and \eves scattering cross sections relevant to the different physics scenarios explored in this work, including both the SM and its extensions.

\subsection{\label{Sec:CEvNS_and_EvES_within_SM}The Standard Model description of \cevns and \eves}

Within the SM, both \cevns and \eves arise from electroweak interactions described by the $SU(2)_L \otimes U(1)_Y$ gauge theory. {At tree level,} the relevant charged current and neutral current interactions of neutrinos, electrons, and quarks with the electroweak gauge bosons are described by the interaction Lagrangian~\cite{Weinberg:1967tq, Salam:1968rm},
\begin{align}\label{eq:SM-CS}
\mathscr{L}_{\rm SM}\supset
&-\frac{g}{2\sqrt{2}}\,\bar{\nu}_e\gamma^{\mu}(1-\gamma^5)e\,W_\mu^+ -\frac{g}{4\cos{\theta_W}}\,\bar{\nu}_\alpha\gamma^{\mu}(1-\gamma^5)\nu_\alpha\,Z_\mu \nonumber\\
&-\frac{g}{2\cos{\theta_W}}\sum_{f=u,d,e}
\bar{f}\gamma^\mu\left[g_{f}^{V\,\mathrm{(SM)}}-g_{f}^{A\,\mathrm{(SM)}}\gamma^5\right]f\,Z_\mu + \mathrm{~H.c.}\,,
\end{align}
where $g$ denotes the $SU(2)_L$ gauge coupling, while $\theta_W$ is the weak mixing angle. Here, $g_{f}^{V\,\mathrm{(SM)}}$ and $g_{f}^{A\,\mathrm{(SM)}}$ denote the tree level vector and axial vector couplings of the fermion $f=\{u,d,e\}$ to the $Z$ boson, whose values are
\begin{equation}
\label{equn:SM_fermion_Couplings}
\begin{aligned}
g_{u}^{V\,\mathrm{(SM)}} &= \frac{1}{2}-\frac{4}{3}\sin^2\theta_W,
&\quad
g_{d}^{V\,\mathrm{(SM)}} &= -\frac{1}{2}+\frac{2}{3}\sin^2\theta_W,
&\quad
g_{e}^{V\,\mathrm{(SM)}} &= -\frac{1}{2}+2\sin^2\theta_W,\\
g_{u}^{A\,\mathrm{(SM)}} &= \frac{1}{2},
&\quad
g_{d}^{A\,\mathrm{(SM)}} &= -\frac{1}{2},
&\quad
g_{e}^{A\,\mathrm{(SM)}} &= -\frac{1}{2}\,.
\end{aligned}
\end{equation}
Using these interaction terms, the SM differential cross sections for \cevns and \eves are discussed below.\\[0.1cm]

\noindent\textbf{CE\textnu NS:} In the SM, \cevns proceeds through a neutral current interaction mediated by the $Z$ boson{, which is flavor independent at tree level}~\cite{Tomalak:2020zfh}. For neutrino energies well below the electroweak scale ($E_\nu \ll M_Z$), integrating out the heavy $Z$ boson [see Eq.~\eqref{eq:SM-CS}] gives rise to the following effective four fermion interaction between neutrinos and quarks. Adopting the SM tree level relation for the electroweak $\rho$ parameter, $\rho=M_W^2/(M_Z^2\cos^2\theta_W)=1$, the corresponding effective interaction Lagrangian takes the form~\cite{Barranco:2005yy}
\begin{equation}
\label{equn:CEvNS_SM_Lagrangian}
\mathscr{L}_\mathrm{SM}^{\nu q} \subset -\frac{G_F}{\sqrt{2}}
\sum_{\substack{q=u,d\\ \alpha=e,\mu,\tau}}
\left[\bar{\nu}_\alpha \gamma^\mu (1-\gamma^5)\nu_\alpha\right]
\left[\bar{q}\gamma_\mu
\left(g_q^{V\,\mathrm{(SM)}}-g_q^{A\,\mathrm{(SM)}}\gamma^5\right)q\right]\,.
\end{equation}
Here, $G_F=g^2/(4\sqrt{2}M_W^2)$ is the Fermi constant, ${\alpha\in\{e,\mu,\tau\}}$ labels the neutrino flavor, and ${q\in\{u,d\}}$ denotes the first generation light quark fields. The quantities $g_q^{V\,\mathrm{(SM)}}$ and $g_q^{A\,\mathrm{(SM)}}$ are the tree-level vector and axial-vector couplings of the quarks, $q$, to the {$Z$ boson, given} in Eq.~\eqref{equn:SM_fermion_Couplings}.
 Vector couplings are explicitly dependent on the weak mixing angle $\theta_W$, a fundamental parameter of electroweak theory~\cite{Weinberg:1967tq, Salam:1968rm}. Its value has been measured with high precision in collider experiments operating at the $Z$ boson pole, yielding $\sin^2\theta_W(M_Z){\equiv\hat{s}_Z^2}=0.23121\pm0.00004$~\cite{ParticleDataGroup:2022pth}, while the corresponding low energy value obtained through renormalization group evolution in the $\overline{\mathrm{MS}}$ scheme is $\sin^2\theta_W(\qtransfer \rightarrow 0){\equiv\hat{s}_0^2}=0.23857\pm0.00005$~\cite{ParticleDataGroup:2022pth, Erler:2019hds}. In contrast to the $Z$ pole determination, the low-energy value of the weak mixing angle remains comparatively less constrained by existing experimental measurements.

{In CE\textnu NS, the quark couplings enter through the effective neutrino-proton and neutrino-neutron vector couplings, which at tree level read $g_V^p=2g_u^{V\,\mathrm{(SM)}}+g_d^{V\,\mathrm{(SM)}}=\frac{1}{2}-2\sin^2\theta_W$ and $g_V^n=g_u^{V\,\mathrm{(SM)}}+2g_d^{V\,\mathrm{(SM)}}=-\frac{1}{2}$. Beyond tree level, these couplings receive electroweak radiative corrections, which we include in the $\overline{\rm MS}$ scheme following Refs.~\cite{Erler:2013xha, ParticleDataGroup:2022pth, Tomalak:2020zfh, AtzoriCorona:2024rtv},}
{\begin{subequations}
\label{eq:gV_RC}
\begin{equation}
\label{eq:gVp_RC}
\begin{split}
g_V^p(\nu_\alpha) = {} & \rho_{\rm NC}\left(\frac{1}{2}-2\sin^2\theta_W\right)+2~\WWcross_{WW}+\square_{WW}-\frac{\sqrt{2}\pi\alpha_{\rm EM}}{3G_F}\langle r^2_{\nu_\alpha}\rangle_{\rm SM}
\\[2pt] &
+\rho_{\rm NC}\left(2\boxtimes_{ZZ}^{uL}+\boxtimes_{ZZ}^{dL}-2\boxtimes_{ZZ}^{uR}-\boxtimes_{ZZ}^{dR}\right)\,,
\end{split}
\end{equation}
\begin{equation}
\label{eq:gVn_RC}
g_V^n = -\frac{\rho_{\rm NC}}{2}+2\square_{WW}+\WWcross_{WW}+\rho_{\rm NC}\left(2\boxtimes_{ZZ}^{dL}+\boxtimes_{ZZ}^{uL}-2\boxtimes_{ZZ}^{dR}-\boxtimes_{ZZ}^{uR}\right)\,,
\end{equation}
\end{subequations}}
{where $\alpha_{\rm EM}\equiv\alpha_{\rm EM}(\qtransfer\rightarrow0)\approx 1/137$ is the electromagnetic coupling in the low momentum transfer limit, i.e. the fine structure constant. $\rho_{\rm NC}=1.00063$ is the low energy neutral current $\rho$ parameter, defined as the ratio of the effective neutral current and charged current four fermion couplings, which deviates from its tree level value $\rho=1$ due to radiative corrections~\cite{Erler:2013xha}. $\square_{WW}$ and $\WWcross_{WW}$ denote the contributions of the $WW$ box and crossed box diagrams, respectively, while $\boxtimes_{ZZ}^{fX}$ ($X=L,R$) denote the contributions of the  $ZZ$ box diagrams,}
{\begin{equation}
\label{eq:box_RC}
\begin{aligned}
\square_{WW}=&-\frac{\hat\alpha_Z}{2\pi\hat s_Z^2}\left[1-\frac{\hat\alpha_s(M_W)}{2\pi}\right],\qquad
\WWcross_{WW}=\frac{\hat\alpha_Z}{8\pi\hat s_Z^2}\left[1+\frac{\hat\alpha_s(M_W)}{\pi}\right],\\
&\boxtimes_{ZZ}^{fX}=-\frac{3\hat\alpha_Z}{8\pi\hat s_Z^2\hat c_Z^2}\left(g_f^{X\,\mathrm{(SM)}}\right)^2\left[1-\frac{\hat\alpha_s(M_Z)}{\pi}\right]\,.
\end{aligned}
\end{equation}}
{Here, $\hat\alpha_Z\equiv\alpha_{\rm EM}(M_Z)\approx1/127$ is the $\overline{\rm MS}$ electromagnetic coupling at the $Z$ pole, $\hat\alpha_s(M_W)$ and $\hat\alpha_s(M_Z)$ are the $\overline{\rm MS}$ strong coupling evaluated at the $W$ and $Z$ boson masses, respectively, and $\hat c_Z^2=1-\hat s_Z^2$. Furthermore, $g_f^{L\,\mathrm{(SM)}}=T_3^f-Q_f\hat s_Z^2$ and $g_f^{R\,\mathrm{(SM)}}=-Q_f\hat s_Z^2$ are the tree level chiral neutral current couplings of the fermion $f$, with $T_3^f$ and $Q_f$ being the third component of weak isospin and the electric charge of the fermion $f$, respectively. The term proportional to $\langle r^2_{\nu_\alpha}\rangle_{\rm SM}$ in Eq.~\eqref{eq:gVp_RC} arises from the SM neutrino charge radius~\cite{Bernabeu:2000hf, Bernabeu:2002nw},}
{\begin{equation}
\label{eq:CR_SM}
\langle r^2_{\nu_\alpha}\rangle_{\rm SM}=-\frac{G_F}{2\sqrt{2}\pi^2}\left[3-2\ln\left(\frac{m_\alpha^2}{M_W^2}\right)\right],
\end{equation}}
{where $m_\alpha$ is the mass of the charged lepton of flavor $\alpha$, and constitutes the only flavor dependent contribution. {Including all the above contributions, one obtains for $\sin^2\theta_W=\hat s_0^2$ the radiatively corrected couplings $g_V^p(\nu_e)\approx0.0382$, $g_V^p(\nu_\mu)\approx0.0300$, $g_V^p(\nu_\tau)\approx0.0256$, and $g_V^n\approx-0.5117$, while the corresponding tree level values are $0.0229$ and $-0.5$.}
Throughout this work, $\sin^2\theta_W$ in Eq.~\eqref{eq:gVp_RC} is fixed to $\hat{s}_0^2$, except when it is extracted from the data, in which case it is treated as a free parameter. In all cases, $\rho_{\rm NC}$, the box contributions (evaluated with $\hat{s}_Z^2$), and the SM charge radius term are fixed to their SM values, so that the extracted $\sin^2\theta_W$ corresponds to the $\overline{\rm MS}$ weak mixing angle at low momentum transfer.} Since \cevns occurs at low momentum transfer, it {therefore} provides a sensitive probe of the weak mixing angle in this regime{, directly comparable with its SM running,} and hence serves as an important test of the SM.

For CE\textnu NS, the quark level effective interaction can be recast in terms of the nuclear current. Since the axial vector contribution vanishes for even-even nuclei and is strongly suppressed for odd-even and odd-odd heavy nuclei due to their small net nuclear spin~\cite{Barranco:2005yy}, only the dominant vector interaction is retained. The resulting effective Lagrangian can be written as
\begin{equation} \label{equn:CEvNS_Nuclear_Level_SM_Lagrangian} \mathscr{L}_\mathrm{SM}^{\nu \mathcal{N}} \subset -\frac{G_F}{\sqrt{2}} \sum_{\substack{\alpha=e,\mu,\tau}} {Q_W^V}\left[\bar{\nu}_\alpha \gamma^\mu (1-\gamma^5)\nu_\alpha\right] \left[\bar{\mathcal{N}}\gamma_\mu \mathcal{N}\right]\,. \end{equation}
The {flavor dependent} effective vector weak charge of the nucleus is given by
\begin{eqnarray} Q_{W}^V = \mathbb{Z}\mathcal{F}_p(\qtransfer^2)\,{g_V^p(\nu_\alpha)} + \mathbb{N}\mathcal{F}_n(\qtransfer^2)\,{g_V^n} \,, \label{eq:weak_charge_SM} \end{eqnarray}
where $\mathbb{Z}$ and $\mathbb{N}$ denote the numbers of protons and neutrons in the target nucleus, respectively, while nuclear physics effects are incorporated through proton and neutron nuclear form factors, $\mathcal{F}_p(\qtransfer^2)$ and $\mathcal{F}_n(\qtransfer^2)$. These quantities account for the finite spatial distribution of nucleons inside the nucleus and become increasingly important as the momentum transfer increases. Their impact is particularly significant for the COHERENT Ge detector, where the momentum transfer is sufficiently large that the loss of coherence cannot be neglected. In contrast, for low energy reactor experiments such as TEXONO, the momentum transfer is much smaller, and the resulting loss of coherence has a negligible effect on the predicted event rate~\footnote{Note that, in our numerical computation, we have considered the effect of  nuclear form factors  for both the experiments.
}. In this work, we employ the effective Klein-Nystrand (KN) parametrization\footnote{We have explicitly verified that adopting other nuclear form factor 
parametrizations commonly employed in the literature, such as the Helm form 
factor~\cite{Helm:1956zz}, leads to no appreciable change in the resulting 
constraints.}~\cite{Klein:1999qj},
\begin{equation}
\mathcal{F}_{p,n}(\qtransfer^2)=3\,
\frac{j_1(\qtransfer R_A)}{\qtransfer R_A}
\left(\frac{1}{1+\qtransfer^2a_k^2}\right)\,,
\end{equation}
with $\qtransfer=\sqrt{2m_{\mathcal{N}}T_{\mathcal{N}}}$ being the three momentum transfer, $j_1$ the spherical Bessel function of the first order, and $a_k=0.7~\mathrm{fm}$ the Yukawa diffuseness parameter. The diffraction radius is related to the proton and neutron rms radii through
\begin{equation}
R_A^2=\frac{5}{3}\langle R_{p,n}\rangle^2-10a_k^2\,,
\end{equation}
where we take $\langle R_p\rangle=4.078~\mathrm{fm}$~\cite{Wang:2024ste, Mann:1973ais, Angeli:2013epw} and $\langle R_n\rangle=4.099~\mathrm{fm}$~\cite{DeRomeri:2026dac} throughout this work.

Using the effective interaction in Eq.~\eqref{equn:CEvNS_Nuclear_Level_SM_Lagrangian}, the SM differential \cevns cross section with respect to the nuclear recoil energy is given by~\cite{Freedman:1973yd, Drukier:1984vhf}
\begin{equation}
\frac{\diff\sigma_{\rm SM}^{\nu\mathcal{N}}}{\diff T_{\mathcal{N}}}
=
\frac{G_F^2m_\mathcal{N}}{\pi}\,
\left(Q_W^V\right)^2
\left[
1
-\frac{m_{\mathcal{N}}T_{\mathcal{N}}}{2E_\nu^2}
\right]\,,
\end{equation}
where $E_\nu$ is the incident neutrino energy, $m_{\mathcal{N}}$ is the target nucleus mass, and $T_{\mathcal{N}}$ denotes the nuclear recoil energy.\\[0.1cm]

\noindent\textbf{E\textnu ES:} In the SM, \eves receives contributions from both charged current and neutral current interactions for electron neutrinos~\cite{Kayser:1979mj}, whereas only the neutral current interaction contributes for muon and tau neutrinos. At neutrino energies well below the electroweak scale ($E_\nu\ll M_{W,Z}$), integrating out the heavy gauge boson  [see Eq.~\eqref{eq:SM-CS}]  leads to the effective four fermion interaction
\begin{equation}
\label{eq:EvES_SM_Lagrangian}
\begin{aligned}
\mathscr{L}_{\rm SM}^{\nu_\alpha e}\subset
-\frac{G_F}{\sqrt{2}}
\Big[
&\left(\bar{\nu}_e\gamma^\mu(1-\gamma^5)e\right)
\left(\bar{e}\gamma_\mu(1-\gamma^5)\nu_e\right)
\\
&+
\sum_{\alpha=e,\mu,\tau}\left(\bar{\nu}_\alpha\gamma^\mu(1-\gamma^5)\nu_\alpha\right)
\left(\bar{e}\gamma_\mu
\left(g_e^{V\,\mathrm{(SM)}}-g_e^{A\,\mathrm{(SM)}}\gamma^5\right)e
\right)
\Big]\,,
\end{aligned}
\end{equation}
where the first term corresponds to the charged current interaction, which contributes only to $\nu_e$-electron scattering, while the second term represents the neutral current interaction common to all neutrino flavors. The tree level vector and axial vector couplings of the electron, $e$, to the SM neutral weak gauge boson, $Z$, are given in Eq.~\eqref{equn:SM_fermion_Couplings}.

Using the effective interaction in Eq.~(\ref{eq:EvES_SM_Lagrangian}), the SM differential \eves cross section with respect to the electron recoil energy, $T_e$, reads as~\cite{Giunti:2007ry}
\begin{equation}
\label{EQ:EvES_SM_xSec}
\frac{\diff\sigma_{\rm SM}^{\nu_\alpha e}}{\diff T_{e}}
=
\frac{G_F^2m_e}{2\pi}
\left[
(g_V+g_A)^2
+(g_V-g_A)^2
\left(1-\frac{T_{e}}{E_\nu}\right)^2
-(g_V^2-g_A^2)
\frac{m_eT_{e}}{E_\nu^2}
\right],
\end{equation}
where{, including the electroweak radiative corrections~\cite{Erler:2013xha},}
{\begin{equation}
\label{eq:gVgA_e_RC}
\begin{aligned}
g_V &= \rho_{\rm NC}\left(-\frac{1}{2}+2\sin^2\theta_W\right)+\square_{WW}+\rho_{\rm NC}\left(\boxtimes_{ZZ}^{eL}-\boxtimes_{ZZ}^{eR}\right)+\frac{\sqrt{2}\pi\alpha_{\rm EM}}{3G_F}\langle r^2_{\nu_\alpha}\rangle_{\rm SM}+\delta_{\alpha e}\,,\\
g_A &= -\frac{\rho_{\rm NC}}{2}+\square_{WW}+\rho_{\rm NC}\left(\boxtimes_{ZZ}^{eL}+\boxtimes_{ZZ}^{eR}\right)+\delta_{\alpha e}\,,
\end{aligned}
\end{equation}}
{where the Kronecker delta $\delta_{\alpha e}$ equals} unity for $\nu_e$-electron scattering and vanishes otherwise, thereby accounting for the additional charged current contribution in $\nu_e$-electron scattering. {The box contributions are given by Eq.~\eqref{eq:box_RC} with $f=e$ and without the QCD correction factors, while the SM charge radius term enters with opposite sign with respect to Eq.~\eqref{eq:gVp_RC}, owing to the opposite electric charge of the electron. {At tree level, Eq.~\eqref{eq:gVgA_e_RC} reduces to $g_V=g_e^{V\,\mathrm{(SM)}}+\delta_{\alpha e}$ and $g_A=g_e^{A\,\mathrm{(SM)}}+\delta_{\alpha e}$, while including the radiative corrections one obtains $g_V(\nu_e)\approx0.9520$, $g_V(\nu_\mu)\approx-0.0398$, $g_V(\nu_\tau)\approx-0.0354$, and $g_A(\nu_e)\approx0.4936$, $g_A(\nu_\mu)\approx-0.5064$, $g_A(\nu_\tau)\approx-0.5064$.}} For antineutrino electron scattering, the corresponding differential cross section is obtained by replacing $g_A\rightarrow-g_A$ in the above expression.

\subsection{\label{Sec:NGI_theory}Neutrino generalized interactions}

Although the SM has been remarkably successful in describing a wide range of particle physics phenomena, nevertheless, the observation of neutrino oscillations and the existence of dark matter provide compelling evidence for physics beyond the SM~\cite{Prajapati:2026tfv, Batra:2026tzz}. Motivated by these open questions, it is natural to consider the possibility of additional interactions mediated by new light particles. In a model independent framework, such interactions can be described by NGIs, which parametrize the most general Lorentz invariant bilinear interactions between neutrinos and first generation quarks or electrons. The corresponding effective interaction Lagrangian is given by~\cite{Lindner:2016wff, AristizabalSierra:2018eqm, Flores:2021kzl}
\begin{equation}
\label{Eq:NGI_Lagrangian}
\mathscr{L}_{\rm NGI}\supset
-\frac{G_F}{\sqrt{2}}
\sum_{\substack{X=S,P,V,A,T\\f=u,d,e\\ \alpha=e,\mu,\tau}}
\varepsilon_{\nu f}^{X}
\left[\bar{\nu}_\alpha\Gamma^{X}\left(1-\gamma^5\right)\nu_\alpha\right]
\left[\bar{f}\Gamma_{X}f\right]\,,
\end{equation}
where $f=\{u,d,e\}$ denotes the first generation SM fermions, $\varepsilon_{\nu f}^{X}$ represents the effective NGI coupling, and
\[
\Gamma_X\equiv
\left\{
\mathbb{I},
i\gamma^5,
\gamma^\mu,
\gamma^\mu\gamma^5,
\sigma^{\mu\nu}
\right\},
\]
corresponding to the scalar ($S$), pseudoscalar ($P$), vector ($V$), axial vector ($A$), and tensor ($T$) Lorentz structures, respectively.

The present analysis employs Ge based detectors, for which the dominant isotopes are even-even nuclei with net spin zero. As a result, the spin dependent pseudoscalar, axial vector, and tensor interactions vanish completely for these isotopes~\cite{Chattaraj:2025rtj}. Although the odd neutron isotope $^{73}$Ge possesses nonzero nuclear spin, its natural abundance ($\sim7.75\%$) is sufficiently small that the corresponding spin dependent contributions remain negligible~\cite{Chattaraj:2025rtj}. Therefore, throughout this work, we restrict our analysis to the spin independent vector and scalar interactions.\\[0.1cm]

\noindent\textbf{Vector interaction:} We first consider the vector NGI and adopt the gauged $U(1)_{B\!-\!L}$ extension of the SM as a representative benchmark realization. This model introduces a neutral vector mediator, $Z'$, associated with the $U(1)_{B\!-\!L}$ gauge symmetry. The corresponding $B\!-\!L$ charges are generation independent and are fixed by anomaly cancellation. The interaction Lagrangian relevant for the \cevns and \eves processes is
\begin{widetext}
\begin{equation}
\mathscr{L}_{B\!-\!L}\supset
\textsl{g}_{B\!-\!L}
\left[
Q_{B\!-\!L}^{q}\bar{q}\gamma^\mu q
+
Q_{B\!-\!L}^{e}\bar{e}\gamma^\mu e
+
Q_{B\!-\!L}^{\nu}\bar{\nu}_\alpha\gamma^\mu\left(\frac{1-\gamma^5}{2}\right)\nu_\alpha
\right]
Z'_\mu
+\frac12M_{Z'}^2Z'_\mu Z'^\mu\,,
\end{equation}
\end{widetext}
where $\textsl{g}_{B\!-\!L}$ is the gauge coupling associated with the $U(1)_{B\!-\!L}$ symmetry, while $M_{Z'}$ denotes the mass of the new vector mediator. The anomaly free $U(1)_{B\!-\!L}$ charge assignments are given by
\begin{equation}
Q_{B\!-\!L}^{q}=\frac13,
\qquad
Q_{B\!-\!L}^{e}=Q_{B\!-\!L}^{\nu}=-1\,.
\end{equation}
Since the $Z'$ interaction has the same Lorentz structure as the SM neutral current interaction, it interferes coherently with the SM amplitude. For \cevns the corresponding effective nuclear charge is~\cite{Bertuzzo:2021opb, Majumdar:2024dms}
\begin{equation}
Q_{Z'}
=
Q_{B\!-\!L}^{\nu}
\left[
\mathbb{Z}\mathcal{F}_p(\qtransfer^2)\left(2Q_{B\!-\!L}^{u}+Q_{B\!-\!L}^{d}\right)
+
\mathbb{N}\mathcal{F}_n(\qtransfer^2)\left(Q_{B\!-\!L}^{u}+2Q_{B\!-\!L}^{d}\right)
\right]\,.
\end{equation}
The resulting differential \cevns cross section in the presence of the $Z'$ mediator becomes~\cite{Bertuzzo:2021opb, Majumdar:2024dms}
\begin{widetext}
\begin{equation}
\frac{\diff\sigma_{\rm SM+Z'}^{\nu\mathcal{N}}}
{\diff T_{\mathcal{N}}}
=
\left[
1+
\frac{
Q_{Z'}\textsl{g}_{B\!-\!L}^2
}
{
\sqrt2\,G_FQ_W^V
\left(M_{Z'}^2+2m_{\mathcal N}T_{\mathcal N}\right)
}
\right]^2
\frac{\diff\sigma_{\rm SM}^{\nu\mathcal{N}}}
{\diff T_{\mathcal{N}}}\,.
\end{equation}
\end{widetext}
For E\textnu ES, the contribution of the $B\!-\!L$ vector mediator can be incorporated into the SM differential cross section in Eq.~\eqref{EQ:EvES_SM_xSec} through the effective replacement~\cite{Lindner:2018kjo}
\begin{equation}
g_V
\rightarrow
g_V'
=
g_V
+
\frac{
\textsl{g}_{B\!-\!L}^2
Q_{B\!-\!L}^{\nu}
Q_{B\!-\!L}^{e}
}
{
\sqrt2\,G_F
\left(M_{Z'}^2+2m_eT_e\right)
}
=
g_V
+
\frac{
\textsl{g}_{B\!-\!L}^2
}
{
\sqrt2\,G_F
\left(M_{Z'}^2+2m_eT_e\right)
}\,.
\end{equation}

\noindent\textbf{Scalar interaction:} We next consider a scalar NGI and adopt a minimal extension of the SM containing a $CP$ even real scalar mediator, $\phi$, with mass $M_\phi$. The interaction Lagrangian relevant for the \cevns and \eves processes is given by
\begin{equation}
\mathscr{L}_{\phi}\supset
\left[
g_\phi^{q}\bar{q}q
+
g_\phi^{e}\bar{e}e
+
g_\phi^{\nu}\bar{\nu}_\alpha\left(\frac{1-\gamma^5}{2}\right)\nu_\alpha
\right]\phi
-\frac12M_\phi^2\phi^2\,.
\end{equation}
Unlike the vector interaction, the scalar interaction has a different Lorentz structure from the SM weak interaction and therefore does not interfere with the SM amplitude. Consequently, the scalar contribution enters additively to both the SM \cevns and \eves differential cross sections.
For CE\textnu NS, the scalar mediated differential cross section is~\cite{Farzan:2018gtr}
\begin{equation}
\frac{\diff\sigma_{\phi}^{\nu\mathcal{N}}}
{\diff T_{\mathcal N}}
=
\frac{
m_{\mathcal N}^{2}
T_{\mathcal N}
Q_{\phi}^{2}
}
{
4\pi
E_{\nu}^{2}
\left(
M_{\phi}^{2}
+
2m_{\mathcal N}T_{\mathcal N}
\right)^{2}
}\,,
\end{equation}
where the effective scalar-nuclear coupling is
\begin{equation}
Q_\phi=
g_{\phi}^{\nu}
\left[
\mathbb{Z}\mathcal{F}_p(\qtransfer^2)
\sum_{q=u,d}
g_{\phi}^{q}
\frac{m_p}{m_q}
f_{T_q}^{p}
+
\mathbb{N}\mathcal{F}_n(\qtransfer^2)
\sum_{q=u,d}
g_{\phi}^{q}
\frac{m_n}{m_q}
f_{T_q}^{n}
\right]\,.
\end{equation}
Here, $m_p$, $m_n$, and $m_q$ denote the proton, neutron, and quark masses, respectively, while $f_{T_q}^{p,n}$ are the scalar-nucleon form factors. The numerical values of these quantities adopted in this work are~\cite{Freeman:2012ry, Alexandrou:2014sha, FlavourLatticeAveragingGroup:2019iem, DelNobile:2021wmp}
\begin{equation}
f_{T_u}^{p}=0.026,\qquad
f_{T_d}^{p}=0.038,\qquad
f_{T_u}^{n}=0.018,\qquad
f_{T_d}^{n}=0.056.
\end{equation}
For E\textnu ES, the scalar contribution to the differential cross section is given by~\cite{Link:2019pbm}
\begin{equation}
\frac{\diff\sigma_{\phi}^{\nu_\alpha e}}
{\diff T_e}
=
\frac{
m_e^2T_e
\left(
g_{\phi}^{\nu}
g_{\phi}^{e}
\right)^2
}
{
4\pi
E_\nu^2
\left(
M_\phi^2
+
2m_eT_e
\right)^2
}\,.
\end{equation}

\subsection{\label{Sec:v_EM_properties}Neutrino electromagnetic properties}

The discovery of neutrino oscillations~\cite{McDonald:2016ixn, Kajita:2016cak}, providing compelling evidence for nonzero neutrino masses~\cite{Pontecorvo:1957cp, Maki:1962mu}, strongly motivates the investigation of possible electromagnetic (EM) interactions of neutrinos~\cite{Schechter:1981hw, Nieves:1981zt, Kayser:1982br, Shrock:1982sc}, which are absent for massless neutrinos in the SM. Such interactions arise naturally in many extensions of the SM and constitute sensitive probes of new physics. In a model independent framework, the effective interaction between neutrinos and the EM field can be written as
\begin{equation}
\mathscr{L}_{\nu_{\rm EM}}
\supset
\bar{\nu}_\alpha
\Gamma^\mu(\mathbf{q})
\left(\frac{1-\gamma^5}{2}\right)
\nu_\alpha
A_\mu\,,
\end{equation}
where $A_\mu$ denotes the EM gauge field and $\Gamma^\mu(\mathbf{q})$ is the most general Lorentz and EM gauge invariant neutrino-photon vertex, parametrized as~\cite{Vogel:1989iv, Nowakowski:2004cv, Giunti:2014ixa, Giunti:2024gec}
\begin{equation}
\Gamma^\mu(\mathbf{q})=
\mathcal{F}_Q(\mathbf{q}^2)\gamma^\mu
-i\mathcal{F}_M(\mathbf{q}^2)\sigma^{\mu\nu}\mathbf{q}_\nu
+\mathcal{F}_E(\mathbf{q}^2)\sigma^{\mu\nu}\mathbf{q}_\nu\gamma^5
+\mathcal{F}_A(\mathbf{q}^2)
\left(
\mathbf{q}^\nu\mathbf{q}_\nu\gamma^\mu
-
\mathbf{q}^\mu\slashed{\mathbf{q}}
\right)\gamma^5\,.
\end{equation}
Here, $\mathbf{q}^\mu$ denotes the four momentum transfer carried by the photon, and $\mathbf{q}^2\equiv\mathbf{q}^\mu\mathbf{q}_\mu$ is the corresponding Lorentz invariant. Since the EM form factors are Lorentz invariant quantities, they depend only on $\mathbf{q}^2$~\cite{Giunti:2014ixa}. For the elastic processes considered in this work the momentum transfer is spacelike, and the four momentum transfer squared is related to the three momentum transfer, $\qtransfer$, by $\mathbf{q}^2=-\qtransfer^2$, with $\qtransfer^2=2m_{\mathcal N}T_{\mathcal N}$ for \cevns and $\qtransfer^2=2m_eT_e$ for E\textnu ES. The form factors $\mathcal{F}_Q$, $\mathcal{F}_M$, $\mathcal{F}_E$, and $\mathcal{F}_A$ correspond to the neutrino charge, magnetic dipole, electric dipole, and anapole interactions, respectively. In the static limit ($\mathbf{q}^2\rightarrow0$), they determine the neutrino millicharge, magnetic moment, electric dipole moment, and anapole moment, while the neutrino charge radius is related to the slope of the charge form factor through~\cite{Giunti:2014ixa}
\begin{equation}
\langle r_\nu^2\rangle
=
\frac{6}{e}
\left.
\frac{\partial\mathcal{F}_Q(\mathbf{q}^2)}
{\partial\mathbf{q}^2}
\right|_{\mathbf{q}^2=0}.
\end{equation}
The effective EM interaction relevant for low energy neutrino scattering can be conveniently expressed in the flavor basis. For short baseline experiments, where neutrino oscillation effects can be neglected, the corresponding effective interaction Lagrangian is given by
\begin{align}
\label{Eq:v_EM_fundamental_Lagrangian}
\mathscr{L}_{\nu_{\rm EM}}\supset
&
\,q_{\nu_\alpha}\,
\bar{\nu}_\alpha\gamma^\mu
\left(\frac{1-\gamma^5}{2}\right)
\nu_\alpha
A_\mu
+\frac{\mu_{\nu_\alpha}^{\rm eff}}{2}\,
\bar{\nu}_\alpha
\sigma^{\mu\nu}
\left(\frac{1-\gamma^5}{2}\right)
\nu_\alpha
F_{\mu\nu}
\nonumber\\
&
+e\,
a_{\nu_\alpha}\,
\bar{\nu}_\alpha
\gamma^\mu\gamma^5
\left(\frac{1-\gamma^5}{2}\right)
\nu_\alpha\,
\partial^\nu F_{\mu\nu}
+e
\frac{\langle r_{\nu_\alpha}^2\rangle}{6}
\bar{\nu}_\alpha
\gamma^\mu
\left(\frac{1-\gamma^5}{2}\right)
\nu_\alpha
\,\partial^\nu F_{\mu\nu}
+\mathrm{H.c.}\,,
\end{align}
where $q_{\nu_\alpha}$, $\mu_{\nu_\alpha}^{\rm eff}$, $a_{\nu_\alpha}$, and $\langle r_{\nu_\alpha}^2\rangle$ denote the neutrino millicharge, effective magnetic moment, anapole moment, and charge radius, respectively, while $e$ is the charge of electron.
For neutrino scattering processes, the outgoing neutrino state is not experimentally observed. Consequently, the magnetic moment contribution is conveniently expressed in terms of an effective magnetic moment, $\mu_{\nu_\alpha}^{\rm eff}$, which incorporates the underlying magnetic and electric dipole moments defined in the neutrino mass basis. For short baseline experiments, it is given by~\cite{Grimus:1997aa, AristizabalSierra:2021fuc}
\begin{equation}
\left(\mu_{\nu_\alpha}^{\rm eff}\right)^2
=
\sum_k
\left|
\sum_j
U_{\alpha j}^{*}
\lambda_{jk}
\right|^2,
\qquad
\lambda_{jk}
=
\mu_{jk}
-
id_{jk},
\end{equation}
where $\mu_{jk}$ and $d_{jk}$ denote the magnetic and electric dipole moments connecting the neutrino mass eigenstates $\nu_j$ and $\nu_k$, respectively, while $U$ is the lepton mixing matrix.

In this work, we investigate the phenomenological implications of the neutrino magnetic moment, millicharge, charge radius, and anapole moment in both \cevns and E\textnu ES. Within the SM, neutrinos do not couple directly to the photon at tree level, while a nonzero but highly suppressed neutrino charge radius and anapole moment are generated at the electroweak one loop level~\cite{Bardeen:1972vi, Lee:1972tnf, Lee:1977tib, Lucio:1983mg, Lucio:1984jn, Dvornikov:2003js, Dvornikov:2004sj}. The minimally extended SM, obtained by introducing three right handed neutrinos, additionally predicts a nonzero but extremely small neutrino magnetic moment arising from electroweak loop effects~\cite{Petcov:1976ff, Marciano:1977wx, Lee:1977tib, Fujikawa:1980yx, Pal:1981rm, Shrock:1982sc, Bernabeu:2000hf, Bernabeu:2002nw, Dvornikov:2003js, Dvornikov:2004sj, Bernabeu:2002pd}. However, many extensions of the SM can significantly enhance these EM properties or even induce a nonzero neutrino millicharge~\cite{Babu:1989wn}. {Consequently, compelling evidence for BSM physics would require either the observation of an EM property that vanishes in the SM, namely a nonzero neutrino millicharge or a magnetic moment, or a measured charge radius (equivalently, anapole moment) incompatible with the SM expectation of Eq.~\eqref{eq:CR_SM}.}

The magnetic moment interaction flips the neutrino helicity and therefore contributes incoherently to the SM cross sections, whereas the millicharge, charge radius, and anapole interactions preserve the neutrino helicity and interfere with the SM amplitude. The corresponding magnetic dipole moment (MDM) contributions to the differential cross sections for \cevns and \eves are given by~\cite{Vogel:1989iv}
\begin{subequations}
\label{Eq:MDM_xsec}
\begin{equation}
\label{eq:CEvNS_MDM_xsec}
\frac{\diff\sigma_{\rm MDM}^{\nu_\alpha\mathcal{N}}}
{\diff T_{\mathcal N}}
=
\frac{\pi\alpha_{\rm EM}^{2}}{m_e^{2}}
\left(
\frac{1}{T_{\mathcal N}}
-
\frac{1}{E_\nu}
\right)
\mathbb{Z}^{2}
\mathcal{F}_{p}^{2}(\qtransfer^{2})
\left(
\frac{\mu_{\nu_\alpha}^{\rm eff}}
{\mu_B}
\right)^2,
\end{equation}
\begin{equation}
\label{eq:EvES_MDM_xsec}
\frac{\diff\sigma_{\rm MDM}^{\nu_\alpha e}}
{\diff T_e}
=
\frac{\pi\alpha_{\rm EM}^{2}}{m_e^{2}}
\left(
\frac{1}{T_e}
-
\frac{1}{E_\nu}
\right)
\left(
\frac{\mu_{\nu_\alpha}^{\rm eff}}
{\mu_B}
\right)^2.
\end{equation}
\end{subequations}
Here, $\mu_B$ denotes the Bohr magneton. Note that, as in Sec.~\ref{Sec:CEvNS_and_EvES_within_SM}, the nuclear form factors $\mathcal{F}_{p,n}$ are evaluated in terms of the three momentum transfer $\qtransfer$, in which the nuclear charge distributions are conventionally parametrized.
The helicity preserving EM interactions modify the SM differential cross sections as~\cite{Giunti:2014ixa, Giunti:2024gec}
\begin{subequations}
\label{Eq:HP_EM_xsec}
\begin{equation}
\label{eq:CEvNS_HP_EM_xsec}
\frac{\diff\sigma_{\rm SM+EM}^{\nu_\alpha\mathcal{N}}}
{\diff T_{\mathcal N}}
=
\frac{G_F^2m_{\mathcal N}}{\pi}
\left[
1-\frac{m_{\mathcal N}T_{\mathcal N}}
{2E_\nu^2}
\right]
\left(
Q_W^V
-
\mathbb{Z}\mathcal{F}_{p}(\qtransfer^{2})Q_\alpha
\right)^2,
\end{equation}
\begin{equation}
\label{eq:EvES_HP_EM_xsec}
\frac{\diff\sigma_{\rm SM+EM}^{\nu_\alpha e}}
{\diff T_e}
=
\frac{G_F^2m_e}{2\pi}
\Bigg[
(g_V+g_A+Q_\alpha)^2
+
(g_V-g_A+Q_\alpha)^2
\left(
1-\frac{T_e}{E_\nu}
\right)^2
-
\left[
(g_V+Q_\alpha)^2-g_A^2
\right]
\frac{m_eT_e}{E_\nu^2}
\Bigg].
\end{equation}
\end{subequations}
where
\begin{equation}
\label{equn:v_millicharge_CR_charge}
Q_\alpha
=
\frac{\sqrt2\pi\alpha_{\rm EM}}{G_F}
\left[
\frac{\langle r_{\nu_\alpha}^2\rangle}{3}
-
2a_{\nu_\alpha}
-
\frac{2}{\qtransfer^2}
\left(
\frac{q_{\nu_\alpha}}{e}
\right)
\right].
\end{equation}
Here, the three momentum transfer is given by $\qtransfer=\sqrt{2m_{\mathcal N}T_{\mathcal N}}$ for \cevns and $\qtransfer=\sqrt{2m_eT_e}$ for E\textnu ES. Note that, throughout this work, when deriving constraints on any one of EM properties (e.g.,  millicharge, charge radius, and anapole moment), we set other properties to zero.
{Moreover, when constraining the neutrino charge radius (anapole moment), the SM charge radius term in Eqs.~\eqref{eq:gVp_RC} and \eqref{eq:gVgA_e_RC} is omitted, so that $\langle r^2_{\nu_\alpha}\rangle$ in Eq.~\eqref{equn:v_millicharge_CR_charge} gives the range of the neutrino charge radius favoured by the data, without fixing its SM contribution \emph{a priori}.}

\subsection{\label{SubSec:Sterile_v_Upscattering}Sterile neutrino upscattering}

Sterile neutrinos or sterile neutral leptons (SNLs) with masses ranging from the keV to MeV scale are well motivated in extensions of the SM, due to their possible connection with neutrino mass generation, dark matter phenomenology, and early universe cosmology \cite{Drewes:2016upu,Abazajian:2017tcc,Boyarsky:2018tvu}.
If  SNLs exist in nature, they can be produced through the upscattering of active neutrinos off nuclei or electrons~\cite{McKeen:2010rx},
leading to characteristic experimental signatures in low energy neutrino experiments. Here, we investigate two representative mechanisms for sterile neutrino upscattering: the transition dipole portal and generalized interactions.\\[0.1cm]

\noindent\textbf{Transition dipole portal:} We first consider the sterile neutrino upscattering scenario mediated by an active-sterile transition magnetic moment, commonly referred to as the transition dipole portal. This interaction can be obtained from the magnetic moment term in the effective electromagnetic interaction of Eq.~\eqref{Eq:v_EM_fundamental_Lagrangian}, replacing the outgoing active neutrino by a massive SNL, $\mathrm{N}$, and the effective magnetic moment, $\mu_{\nu_\alpha}^{\rm eff}$, by the corresponding active-sterile transition magnetic moment, $\mu_{\nu_\alpha \mathrm{N}}^{\rm eff}$. Consequently, an incoming active neutrino can undergo the inelastic upscattering into a heavier sterile neutral lepton through the processes $\nu_\alpha+\mathcal{N}\rightarrow \mathrm{N}+\mathcal{N}$, and $\nu_\alpha+e\rightarrow \mathrm{N}+e$, through photon exchange. Since the transition dipole interaction flips the neutrino helicity and produces a different final state particle, its contribution does not interfere with the SM amplitude and therefore contributes incoherently to the total event rate. The corresponding differential cross sections for \cevns and \eves are given by~\cite{McKeen:2010rx, Chen:2021uuw, Miranda:2021kre}
\begin{subequations}
\label{Eq:sterile_DP_xsec}
\begin{equation}
\label{eq:CEvNS_sterile_DP_xsec}
\begin{split}
\frac{\diff\sigma_{\rm DP}^{\nu_\alpha\mathcal{N}-\mathrm{N}\mathcal{N}}}
{\diff T_{\mathcal N}}
= \frac{\pi\alpha_{\rm EM}^{2}}{m_e^{2}}
\Bigg[ &
\frac{1}{T_{\mathcal N}}
- \frac{1}{E_\nu}
- \frac{m_\mathrm{N}^2}{2E_\nu T_\mathcal{N} m_\mathcal{N}}
    \left(1- \frac{T_\mathcal{N}}{2E_\nu} + \frac{m_\mathcal{N}}{2E_\nu}\right)
\\[2pt] &
    + \frac{m_\mathrm{N}^4\,(T_\mathcal{N}-m_\mathcal{N})}{8E_\nu^2 T_\mathcal{N}^2 m_\mathcal{N}^2}
\Bigg]
\mathbb{Z}^{2}\,
\mathcal{F}_{p}^{2}(\qtransfer^{2})
\left(\frac{\mu_{\nu_\alpha \mathrm{N}}^{\rm eff}}{\mu_B}\right)^{\!2}\,,
\end{split}
\end{equation}
\begin{equation}
\label{eq:EvES_sterile_DP_xsec}
\frac{\diff\sigma_{\rm DP}^{\nu_\alpha e-\mathrm{N}e}}
{\diff T_e}
=
\frac{\pi\alpha_{\rm EM}^{2}}{m_e^{2}}
\left[
\frac{1}{T_e}
-
\frac{1}{E_\nu}
- \frac{m_\mathrm{N}^2}{2E_\nu T_e m_e}
    \left(1- \frac{T_e}{2E_\nu} + \frac{m_e}{2E_\nu}\right)
    + \frac{m_\mathrm{N}^4(T_e-m_e)}{8E_\nu^2 T_e^2 m_e^2}
\right]
\left(
\frac{\mu_{\nu_\alpha \mathrm{N}}^{\rm eff}}
{\mu_B}
\right)^2\,,
\end{equation}
\end{subequations}
respectively.\\[0.1cm]

\noindent\textbf{Upscattering production of sterile neutral leptons via generalized interactions:} We next consider sterile neutrino upscattering mediated by generalized interactions. The corresponding interaction is obtained by replacing the outgoing active neutrino in the effective NGI Lagrangian of Eq.~\eqref{Eq:NGI_Lagrangian} with a massive SNL, $\mathrm{N}$. The effective interaction Lagrangian describing sterile neutrino upscattering via generalized interactions can be written as
\begin{equation}
\mathscr{L}_{\rm SNL}
\supset
\sum_{\substack{X=S,P,V,A,T\\f=u,d,e\\ \alpha=e,\mu,\tau}}
\frac{Q_X^{\,\nu}Q_X^{\,f}}
{\mathbf{q}^\mu\mathbf{q}_\mu-M_X^2}
\left[
\bar{\mathrm{N}}\Gamma^{X}\left(\frac{1-\gamma^5}{2}\right)\nu_\alpha
\right]
\left[
\bar{f}\Gamma_{X}f
\right]\,,
\end{equation}
where $Q_X^{\,\nu}$ and $Q_X^{\,f}$ denote the couplings of the mediator $X$ to the neutrino and the fermion $f$, respectively. However, as discussed in Sec.~\ref{Sec:NGI_theory}, only the spin independent scalar ($S$) and vector ($V$) interactions are considered throughout this work. Consequently, an incoming active neutrino can undergo the inelastic upscattering processes $\nu_\alpha+\mathcal{N}\rightarrow \mathrm{N}+\mathcal{N}$ and $\nu_\alpha+e\rightarrow \mathrm{N}+e$ mediated by either a scalar or a vector boson. The corresponding differential cross sections for \cevns are given by~\cite{Candela:2024ljb}
\begin{subequations}
\label{Eq:sterile_NGI_CEvNS_xsec}
\begin{equation}
\label{eq:CEvNS_sterile_scalar_xsec}
\frac{\diff\sigma_{S}^{\nu\mathcal{N}-\mathrm{N}\mathcal{N}}}
{\diff T_{\mathcal N}}
=
\frac{
m_{\mathcal N}
Q_{S}^{2}
}
{
4\pi
\left(
M_{S}^{2}
+
2m_{\mathcal N}T_{\mathcal N}
\right)^{2}
}
\left(
\frac{m_{\mathcal N}T_{\mathcal N}}
{E_{\nu}^{2}}
+
\frac{m_\mathrm{N}^{2}}
{2E_{\nu}^{2}}
\right),
\end{equation}
\begin{equation}
\label{eq:CEvNS_sterile_vector_xsec}
\frac{\diff\sigma_{V}^{\nu\mathcal{N}-\mathrm{N}\mathcal{N}}}
{\diff T_{\mathcal N}}
=
\frac{
m_{\mathcal N}
Q_{V}^{2}
}
{
2\pi
\left(
M_{V}^{2}
+
2m_{\mathcal N}T_{\mathcal N}
\right)^{2}
}
\left[
\left(
1
-
\frac{m_{\mathcal N}T_{\mathcal N}}
{2E_{\nu}^{2}}
\right)
-
\frac{m_\mathrm{N}^{2}}
{4E_{\nu}^{2}}
\left(
1
+
\frac{2E_{\nu}}
{m_{\mathcal N}}
\right)
\right]\,,
\end{equation}
\end{subequations}
where the effective vector and scalar-nuclear couplings are
\begin{equation}
Q_{V}
=
Q_{V}^{\nu}
\left[
\mathbb{Z}\mathcal{F}_p(\qtransfer^2)
\left(
2Q_{V}^{u}
+
Q_{V}^{d}
\right)
+
\mathbb{N}\mathcal{F}_n(\qtransfer^2)
\left(
Q_{V}^{u}
+
2Q_{V}^{d}
\right)
\right],
\end{equation}
and
\begin{equation}
Q_{S}
=
Q_{S}^{\nu}
\left[
\mathbb{Z}\mathcal{F}_p(\qtransfer^2)
\sum_{q=u,d}
Q_{S}^{q}
\frac{m_p}{m_q}
f_{T_q}^{p}
+
\mathbb{N}\mathcal{F}_n(\qtransfer^2)
\sum_{q=u,d}
Q_{S}^{q}
\frac{m_n}{m_q}
f_{T_q}^{n}
\right]\,.
\end{equation}

Similarly, the corresponding differential cross sections for \eves are~\cite{Candela:2024ljb}
\begin{subequations}
\label{Eq:sterile_NGI_EvES_xsec}
\begin{equation}
\label{eq:EvES_sterile_scalar_xsec}
\frac{\diff\sigma_{S}^{\nu_\alpha e-\mathrm{N}e}}
{\diff T_{e}}
=
\frac{
m_{e}
\left(Q_{S}^{\nu}Q_{S}^{e}\right)^{2}
}
{
4\pi
\left(
M_{S}^{2}
+
2m_{e}T_{e}
\right)^{2}
}
\left(
1+
\frac{T_{e}}
{2m_{e}}
\right)
\left(
\frac{m_{e}T_{e}}
{E_{\nu}^{2}}
+
\frac{m_\mathrm{N}^{2}}
{2E_{\nu}^{2}}
\right),
\end{equation}
\begin{equation}
\label{eq:EvES_sterile_vector_xsec}
\frac{\diff\sigma_{V}^{\nu_\alpha e-\mathrm{N}e}}
{\diff T_{e}}
=
\frac{
m_{e}
\left(Q_{V}^{\nu}Q_{V}^{e}\right)^{2}
}
{
2\pi
\left(
M_{V}^{2}
+
2m_{e}T_{e}
\right)^{2}
}
\left[
\left(
1
-
\frac{m_{e}T_{e}}
{2E_{\nu}^{2}}
-
\frac{T_{e}}
{E_{\nu}}
+
\frac{T_{e}^{2}}
{2E_{\nu}^{2}}
\right)
-
\frac{m_\mathrm{N}^{2}}
{4E_{\nu}^{2}}
\left(
1
+
\frac{2E_{\nu}}
{m_{e}}
-
\frac{T_{e}}
{m_{e}}
\right)
\right]\,,
\end{equation}
\end{subequations}
respectively. Throughout this work, we adopt the universal benchmark
$
Q_X^{u}=Q_X^{d}=Q_X^{e}\equiv Q_X^{f},
$
and express our results in terms of the effective coupling
$
\textsl{g}_{X}=\sqrt{Q_X^{\nu}Q_X^{f}},
$
since the scattering cross sections depend on the neutrino and target couplings only through their product.

\section{\label{Sec:Events}Events simulation and statistical analysis}
In this section, we present the experimental inputs, describe the procedure used to simulate the expected events in each detector, and outline the statistical procedure adopted in the analysis of the COHERENT Ge-mini, and TEXONO datasets.

\subsection{COHERENT Ge-mini}

The COHERENT Ge-mini experiment employs an array of four ICPC HPGe detectors, named as Ge-21, Ge-25, Ge-26, and Ge-28, operated at the SNS to measure \cevns with sub-keV energy thresholds~\cite{COHERENT:2026yje,Bouabid:2025upo}. The fiducial masses of the individual detectors are summarized in Table~\ref{tab:GeMini_detectors}. The total fiducial mass of the detector array is $m_{\rm det}=8.53~{\rm kg}$. To date, the COHERENT Ge-mini experiment has completed two data taking campaigns: the first was conducted between June and August 2023~\cite{COHERENT:2024axu}, while the second took place between February and May 2025~\cite{COHERENT:2026yje}. In the present work, we consider the data from the second campaign owing to its larger exposure and improved statistics.

For a given interaction channel $\xi$, the predicted number of reconstructed nuclear events in the $i$th energy bin is calculated as
\begin{equation}
\label{eq:GeMini_CEvNS_events}
\left[R_i\right]_\xi^{\nu_\alpha\mathcal N}
=
N_{\rm target}
\int_{T_{e,i}}^{T_{e,i+1}}
\diff T_e^{\rm reco}
\int_{T_\mathcal N^{\rm min}}^{T_\mathcal N^{\rm max}}
\diff T_\mathcal N\,
G(T_e^{\rm reco},T_e)
\int_{E_\nu^{\rm min}}^{E_\nu^{\rm max}}
\diff E_\nu\,
\frac{\diff\Phi_{\nu_\alpha}}{\diff E_\nu}
\frac{\diff\sigma_\xi^{\nu_\alpha\mathcal N}}
{\diff T_\mathcal N},
\end{equation}
where $N_{\rm target}
=
m_{\rm det}/m_\mathcal N$
denotes the total number of target nuclei.
The neutrino flux from the SNS at the detector consists of one prompt component originating from pion decay at rest ($\pi$-DAR) and two delayed components arising from muon decay at rest ($\mu$-DAR)~\cite{Michel:1949qe,Bouchiat:1957zz},
\begin{equation}
\begin{aligned}
\frac{\diff\Phi_{\nu_\mu}}
{\diff E_\nu}
&=
\eta\,
\delta\!\left(
E_\nu-
\frac{m_\pi^2-m_\mu^2}{2m_\pi}
\right)
 \quad &(\text{prompt})\, , \\
\frac{\diff\Phi_{\bar\nu_\mu}}
{\diff E_\nu}
&=
\eta
\frac{64E_\nu^2}{m_\mu^3}
\left(
\frac34-\frac{E_\nu}{m_\mu}
\right)
 \quad &(\text{delayed})\, ,\\
\frac{\diff\Phi_{\nu_e}}
{\diff E_\nu}
&=
\eta
\frac{192E_\nu^2}{m_\mu^3}
\left(
\frac12-\frac{E_\nu}{m_\mu}
\right) \quad &(\text{delayed})\, ,
\end{aligned}
\end{equation}
where $m_\pi$ and $m_\mu$ are the masses of the pion and muon, respectively. The SNS flux normalization is
$
\eta
=
rN_{\rm POT}/4\pi L^2
$,
with $r=0.123$ neutrinos per flavor produced per proton on target (POT), $N_{\rm POT}=4.68\times10^{22}$ is the accumulated POT during the data taking period, and $L=19.2~{\rm m}$ is the detector baseline from the SNS. For standard \cevns interactions, the minimum neutrino energy required to produce a nuclear recoil $T_\mathcal N$ is approximated by,
$E_\nu^{\rm min}
\simeq
\sqrt{m_\mathcal NT_\mathcal N/2}$,
whereas for SNL upscattering the kinematic threshold is modified to\footnote{\label{FootNote:Sterile_Mass_Limit_CEvNS}The kinematics of the process also imposes an upper bound on the SNL mass. For a given incident neutrino energy and nuclear recoil energy, the kinematically accessible SNL mass satisfies
$
m_\mathrm{N}^2 \lesssim 2m_\mathcal{N}T_\mathcal{N}
\left(
\sqrt{\frac{2}{m_\mathcal{N}T_\mathcal{N}}}E_\nu-1
\right).
$}
\begin{equation}
E_\nu^{\rm min}
\simeq
\sqrt{\frac{m_\mathcal NT_\mathcal N}{2}}
\left(
1+
\frac{m_\mathrm{N}^2}
{2m_\mathcal NT_\mathcal N}
\right).
\end{equation}
The maximum recoil energy is approximated by
$T_\mathcal N^{\rm max}
\simeq
2(E_\nu^{\rm max})^2/m_\mathcal N$.
The measured observable is the electron equivalent ionization energy, $T_e$, which is related to the nuclear recoil energy through the quenching factor,
$T_e
=
\mathcal Q_f(T_\mathcal N)\,
T_\mathcal N$.
We employ the standard Lindhard quenching model~\cite{Lindhard:1963} with parameter $k=0.157$, as recommended by the COHERENT collaboration~\cite{COHERENT:2026yje}.
Detector resolution is incorporated through a truncated Gaussian response function,
\begin{equation}
\label{Eq:Gaussian_Smearing_Func}
G(T_e^{\rm reco},T_e)
=
\frac{1}
{\sqrt{2\pi}\sigma_e}
\exp
\left[
-
\frac{(T_e^{\rm reco}-T_e)^2}
{2\sigma_e^2}
\right],
\end{equation}
where $T_e^{\rm reco}$ denotes the reconstructed ionization energy. The energy resolution is parametrized as~\cite{Bouabid:2025upo}
\begin{equation}
\sigma_e(T_e)
=
\sqrt{
\sigma_{\rm noise}^2
+
\eta_eF_fT_e
},
\end{equation}
where $\eta_e=2.96~{\rm eV_{ee}}$ is the average energy required to produce an electron-hole pair in germanium at 77 K~\cite{ANTMAN1966272, Wei:2016xbw}, $F_f$ is the detector dependent Fano factor, and the electronic noise resolution is obtained from the pulser full width at half maximum (FWHM) according to
\begin{equation}
\sigma_{\rm noise}
=
\frac{\rm FWHM}
{\sqrt{8\ln2}}.
\end{equation}
The Fano factor and pulser FWHM for different detectors of COHERENT Ge-mini are summarized in Table~\ref{tab:GeMini_detectors}. Since these quantities are not explicitly provided in the official COHERENT Ge-mini data release, we adopt the corresponding values reported in Ref.~\cite{Bouabid:2025upo}.
The nuclear recoil spectrum is evaluated above the physical threshold corresponding to an electron equivalent ionization energy of $T_e^{\rm min}\simeq2.96~{\rm eV_{ee}}$, which is the minimum energy required to produce an electron-hole pair in germanium at 77~K.

\begin{table}[ht!]
\centering
\setlength{\tabcolsep}{14pt} % default is 6pt
\begin{tabular}{lccc}
\hline\hline
Detector &
Fiducial mass (kg) &
Pulser FWHM (eV$_{\rm ee}$) &
Fano factor\\
\hline
Ge-21 & 2.13 & 116 & 0.068 \\
Ge-25 & 2.13 & 145 & 0.064 \\
Ge-26 & 2.13 & 152 & 0.068 \\
Ge-28 & 2.14 & 135 & 0.064 \\
\hline
total & 8.53 & -- & -- \\
\hline\hline
\end{tabular}
\caption{\footnotesize Detector specific parameters of the COHERENT Ge-mini detector array. The pulser FWHM and Fano factors are taken from Ref.~\cite{Bouabid:2025upo}.}
\label{tab:GeMini_detectors}
\end{table}

The calculation of reconstructed electron recoil events follows a similar procedure. For a given interaction channel $\xi$, the predicted number of reconstructed electron recoil events in the $i$th energy bin is given by
\begin{equation}
\label{eq:GeMini_EvES_events}
\left[R_i\right]_\xi^{\nu_\alpha e}
=
N_{\rm target}
\int_{T_{e,i}}^{T_{e,i+1}}
\diff T_e^{\rm reco}
\int_{T_e^{\rm min}}^{T_e^{\rm max}}
\diff T_e\,
G(T_e^{\rm reco},T_e)\,
\mathbb{Z}_{\rm eff}(T_e)
\int_{E_\nu^{\rm min}}^{E_\nu^{\rm max}}
\diff E_\nu\,
\frac{\diff\Phi_{\nu_\alpha}}{\diff E_\nu}
\frac{\diff\sigma_\xi^{\nu_\alpha e}}
{\diff T_e},
\end{equation}
where {the \eves cross sections given in Sec.~\ref{Sec:Theory} are derived in the free electron approximation, that is, assuming free and stationary target electrons, and atomic binding effects are accounted for through} $\mathbb{Z}_{\rm eff}(T_e)$, {the} effective number of atomic electrons available for ionization at recoil energy $T_e$. {This correction is implemented} through the step function approximation proposed in Ref.~\cite{Chen:2016eab}, as
$\mathbb{Z}_{\rm eff}(T_e)
=
\sum_{j=1}^{32}
\Theta(T_e-\mathscr{B}_j)
$,
where $\Theta(x)$ is the Heaviside step function and $\mathscr{B}_j$ denotes the binding energy of the $j$th electron in a germanium atom. The corresponding binding energies are taken from Ref.~\cite{xray_data_booklet}.
{We note that the stepping approximation is an effective description of atomic binding effects, which become more important as the electron recoil energy decreases and approaches the atomic binding scale. Dedicated atomic calculations for germanium based on ab initio many body methods~\cite{Chen:2014ypv, Chen:2015hta, Hsieh:2019hug} show that the stepping approximation agrees well with the full atomic treatment for the SM and magnetic moment cross sections, so that the corresponding constraints are largely unaffected. For a neutrino millicharge, whose cross section grows as $1/\qtransfer^4$, the full atomic treatment predicts a larger ionization rate at low recoil energies, and our limits are therefore conservative. The same applies to the light mediator scenarios, where the relevant quantity is the mediator mass: for $M_X\gtrsim\qtransfer$ the interaction is effectively contactlike and the stepping approximation is adequate.  For sufficiently light mediators the cross section approaches the $1/\qtransfer^4$ behavior of the millicharge case, and thus, our analysis again gives conservative bounds.}

For standard neutrino electron scattering, the minimum neutrino energy required to produce an electron recoil $T_e$ is given by
$
E_\nu^{\rm min}
=
\left(T_e+\sqrt{T_e^2+2m_eT_e}\right)
/2$,
whereas for SNL upscattering the kinematic threshold is modified to\footnote{\label{FootNote:Sterile_Mass_Limit_EvES}Analogous to the \cevns case, the inelastic kinematics impose an upper bound on the SNL mass. For a given incident neutrino energy and electron recoil energy, the kinematically accessible SNL mass satisfies
$
m_\mathrm{N}^2
\le
2m_eT_e
\left(
\frac{2E_\nu}
{T_e+\sqrt{T_e^2+2m_eT_e}}
-1
\right).
$
}
\begin{equation}
E_\nu^{\rm min}
=
\frac{
T_e+\sqrt{T_e^2+2m_eT_e}
}{2}
\left(
1+
\frac{m_\mathrm{N}^2}
{2m_eT_e}
\right).
\end{equation}
The maximum electron recoil energy is given by
$
T_e^{\rm max}
=
2(E_\nu^{\rm max})^2/\left(2E_\nu^{\rm max}+m_e\right)
$.

\begin{figure}
    \centering
    \includegraphics[width=0.75\linewidth]{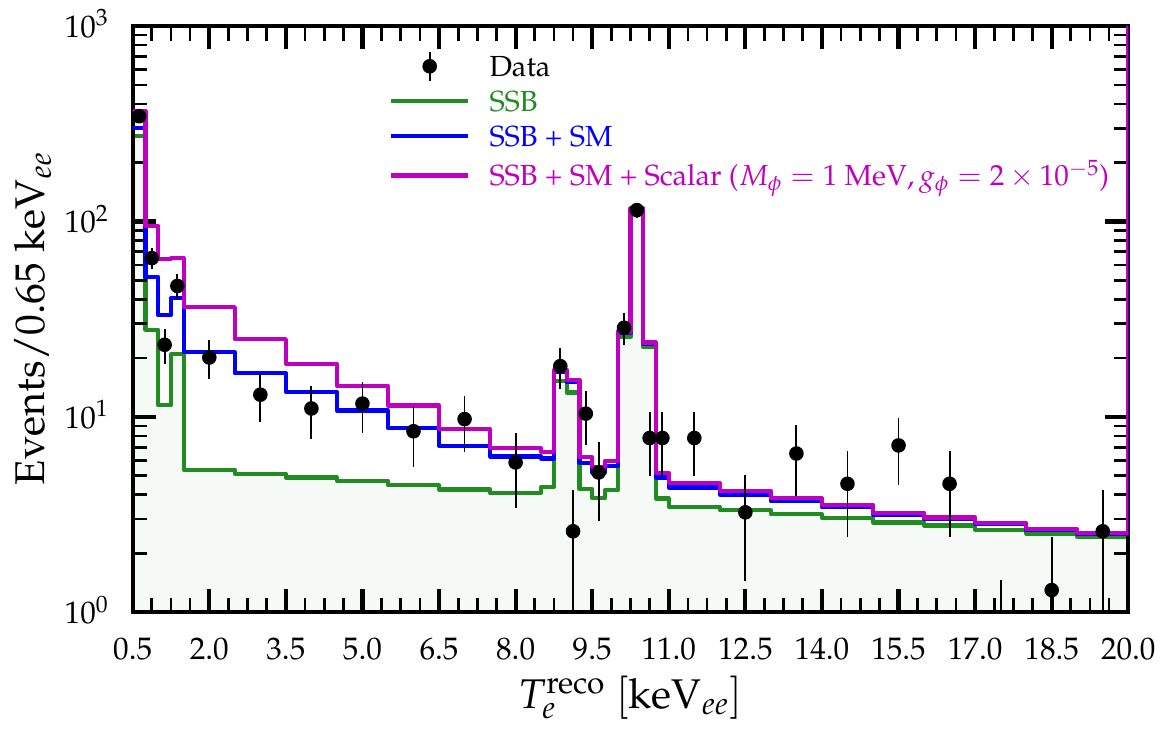}
    \caption{\footnotesize Reconstructed electron equivalent energy spectrum of the COHERENT Ge-mini experiment. The green histogram shows the measured SSB~\cite{COHERENT:2026yje}, the black points represent the observed beam on data~\cite{COHERENT:2026yje}, and the blue (magenta) histogram corresponds to the simulated SM (SM+scalar) event spectrum including both \cevns and \eves contributions.}
    \label{fig:GeMini_spectrum}
\end{figure}

For both nuclear and electron recoil channels, the total event rate is obtained by summing the contributions from all naturally occurring germanium isotopes according to their isotopic abundances. We consider all stable germanium isotopes with mass numbers $\{70,72,73,74,76\}$ and their corresponding natural abundances $\{20.57\%,27.45\%,7.75\%,36.50\%,7.73\%\}$ taken from Ref.~\cite{BerglundWieser:2011}. Finally, the reconstructed event spectrum is obtained by combining the contributions from all isotopes and considering 30 electron equivalent energy bins spanning the interval $0.5$--$20~{\rm keV_{ee}}$, with bin widths as provided in the official COHERENT Ge-mini data release.

Figure~\ref{fig:GeMini_spectrum} shows the reconstructed electron equivalent energy spectrum of the COHERENT Ge-mini experiment as a function of the reconstructed electron equivalent recoil energy, $T_e^{\rm reco}$. The green histogram represents the measured steady state background (SSB)~\cite{COHERENT:2026yje}, while the black data points correspond to the observed beam on events~\cite{COHERENT:2026yje}. The blue and magenta histograms denote the predicted total event spectra, including contributions from both \cevns and E\textnu ES, for the SM and a representative scalar mediated new physics scenario with benchmark parameters $M_\phi=1~\mathrm{MeV}$ and $\textsl{g}_\phi=2\times10^{-5}$, respectively.

To constrain the model parameters, we employ the following Poissonian $\chi^2$ test statistic:
\begin{equation}
\label{eq:chi2}
\chi^2(\boldsymbol{\Theta})
=
2\sum_{i=1}^{30}
\left[
R_i^{\rm th}(\boldsymbol{\Theta})
-
R_i^{\rm exp}
+
R_i^{\rm exp}
\ln
\left(
\frac{R_i^{\rm exp}}
{R_i^{\rm th}(\boldsymbol{\Theta})}
\right)
\right]
+
\left(\frac{\alpha}{\sigma_\alpha}\right)^2
+
\left(\frac{\beta}{\sigma_\beta}\right)^2\,,
\end{equation}
where $R_i^{\rm exp}$ denotes the experimentally observed number of events recorded by the COHERENT collaboration in the $i^\mathrm{th}$ energy bin~\cite{COHERENT:2026yje}, and $\boldsymbol{\Theta}$ denotes the set of model parameters under consideration. The predicted number of events in the $i^\mathrm{th}$ energy bin is given by
\begin{equation}
\label{eq:Rth}
R_i^{\rm th}(\boldsymbol{\Theta})
=
(1+\alpha)
\left[
R_i^{\nu\mathcal N}(\boldsymbol{\Theta})
+
R_i^{\nu e}(\boldsymbol{\Theta})
\right]
+
(1+\beta)
R_i^{\rm SSB}\,,
\end{equation}
where $R_i^{\nu\mathcal N}$ and $R_i^{\nu e}$ denote the predicted \cevns and \eves events, respectively, calculated using Eqs.~\eqref{eq:GeMini_CEvNS_events} and \eqref{eq:GeMini_EvES_events}, while $R_i^{\rm SSB}$ represents the total number of measured steady state background events in the corresponding energy bin taken from Ref.~\cite{COHERENT:2026yje}. The nuisance parameter $\alpha$ accounts for the overall signal normalization systematic uncertainty, obtained by combining in quadrature the uncertainties associated with the neutrino flux normalization {(10\%)}, baseline distance {(0.5\%)}, energy calibration {(0.8\%)}, active detector mass {(2\%)}, nuclear form factor {(0.8\%)}, and germanium quenching factor {(0.7\%)}. The nuisance parameter $\beta$ parametrizes the background normalization systematic uncertainty. The corresponding prior uncertainties are taken to be $\sigma_\alpha=10.3\%$~\cite{COHERENT:2026yje}\footnote{{The quenching factor and the nuclear form factor can in principle also
distort the spectral shape. We have checked this by varying $k$ and
$\langle R_n \rangle$ by amounts that change the total predicted rate by the
corresponding systematic uncertainties quoted above ($0.7\%$ and $0.8\%$, respectively).
The resulting spectral distortions stay well below the per bin statistical uncertainty.
This is consistent with Refs.~\cite{AtzoriCorona:2026wbu,DeRomeri:2026dac}, where these
quantities are marginalized over, and the impact on the extracted results is found to be
marginal. Treating both as normalization uncertainties is therefore adequate here.}} and $\sigma_\beta=1\%$~\cite{COHERENT:2024axu}.
For each point in the model parameter space, $\boldsymbol{\Theta}$, the corresponding $\chi^2(\boldsymbol{\Theta})$ is obtained by minimizing Eq.~\eqref{eq:chi2} with respect to the nuisance parameters $\alpha$ and $\beta$. The resulting profiled $\chi^2$ is then used to derive the allowed and excluded regions in the parameter space.
{Fitting the normalization of the SM CE\textnu NS signal to the data, we obtain a total of $125\pm10$ events at $1\sigma$, in good agreement with our theoretical SM prediction of $127$ events and with the $124^{+14}_{-12}$ events ($1\sigma$) reported by the COHERENT collaboration~\cite{COHERENT:2026yje}.}

{We conclude by noting that our COHERENT Ge-mini analysis is restricted to the reconstructed energy spectrum, which retains the spectral information of the signal. It does not exploit the timing information of the SNS neutrino flux, which provides additional discrimination between the prompt $\nu_\mu$ component and the delayed $\nu_e$ and $\bar\nu_\mu$ components. A two dimensional analysis combining energy and timing would in principle be preferable, but the correlated energy-time distribution is not publicly available~\cite{COHERENT:2026yje}. We leave its inclusion for future work, once the correlated dataset is released.}

\subsection{TEXONO}
The TEXONO experiment employs two p-type ECPCGe detectors, denoted as D70 and D50, operated at the KSNL to search for reactor based \cevns \cite{TEXONO:2026eyr}. The detector specific parameters are summarized in Table~\ref{tab:TEXONO_detectors}. The experiment is located at a distance of $d=28~{\rm m}$ from the reactor core. In the present work, we analyze the combined D70 and D50 datasets, corresponding to total valid Reactor ON (OFF) exposures of $404~(813.7)~{\rm kg\text{-}days}$.

\begin{table}[ht!]
\centering
\setlength{\tabcolsep}{14pt} % default is 6pt
\begin{tabular}{lcc}
\hline\hline
Parameter & D70 & D50 \\
\hline
fiducial mass (g) & 1334 & 472 \\
threshold (eV$_{\rm ee}$) & 200 & 200 \\
pedestal RMS noise, $\sigma_{\rm noise}$ (eV$_{\rm ee}$) & 31.3 & 31.1 \\
reactor ON exposure (kg-days) & 242 & 162 \\
reactor OFF exposure (kg-days) & 559.3 & 254.4 \\
\hline\hline
\end{tabular}
\caption{\footnotesize Detector specific parameters of the TEXONO ECPCGe detectors~\cite{TEXONO:2026eyr}.}
\label{tab:TEXONO_detectors}
\end{table}

The predicted \cevns and \eves event rates are computed following the same formalism as described for the COHERENT Ge-mini experiment in Eqs.~\eqref{eq:GeMini_CEvNS_events} and \eqref{eq:GeMini_EvES_events}, respectively. The primary differences arise from the reactor antineutrino flux and the detector response specific to the TEXONO ECPCGe detectors. The reactor $\bar{\nu}_e$ flux is given by
\begin{equation}
\label{eq:TEXONO_flux}
\frac{\diff\Phi_{\bar{\nu}_e}}{\diff E_\nu}
=
\frac{P}{4\pi d^2\epsilon}
\sum_k
f_k
\frac{\diff N_{\bar{\nu}_e}^{\,k}}
{\diff E_\nu},
\end{equation}
where $P=2.9~\mathrm{GW}$ is the reactor thermal power and $\epsilon=205.24~\mathrm{MeV}$ is the average energy released per fission. The summation includes contributions from the four dominant fissile isotopes, $\{{}^{235}\mathrm{U},{}^{238}\mathrm{U},{}^{239}\mathrm{Pu},{}^{241}\mathrm{Pu}\}$, together with the ${}^{238}\mathrm{U}(n,\gamma){}^{239}\mathrm{U}$ neutron capture component. The normalized antineutrino spectra, $\diff N_{\bar{\nu}_e}^{\,k}/\diff E_\nu$, are taken from the Huber-Mueller model~\cite{Huber:2011wv,Mueller:2011nm} for $E_\nu>2~\mathrm{MeV}$, from Ref.~\cite{Vogel:1989iv} for $E_\nu<2~\mathrm{MeV}$, and from Ref.~\cite{TEXONO:2006xds} for the neutron capture component. Following Ref.~\cite{TEXONO:2006xds}, the total reactor antineutrino spectrum is obtained by summing the individual normalized antineutrino spectra weighted by their corresponding neutrino yields per fission, $f_k$. The neutrino yields per fission are taken to be $f_k=\{3.4,\,0.5,\,1.8,\,0.4,\,1.2\}$, corresponding to $\{{}^{235}\mathrm{U},{}^{238}\mathrm{U},{}^{239}\mathrm{Pu},{}^{241}\mathrm{Pu},{}^{238}\mathrm{U}(n,\gamma){}^{239}\mathrm{U}\}$, respectively~\cite{TEXONO:2006xds}, resulting in a total reactor antineutrino flux of $\Phi_{\bar{\nu}_e}\approx 6.4\times10^{12}~\mathrm{cm^{-2}\,s^{-1}}$ at the detector location~\cite{TEXONO:2026eyr}. The nuclear recoil energy is converted to electron equivalent energy using the standard Lindhard quenching model, as described for the COHERENT Ge-mini analysis, with the Lindhard parameter fixed to $k=0.162$ following the recommendation of the TEXONO collaboration~\cite{TEXONO:2026eyr}. The detector response for TEXONO is modeled using the same Gaussian smearing function given in Eq.~\eqref{Eq:Gaussian_Smearing_Func}. However, the detector resolution is evaluated using the pedestal RMS noise values listed in Table~\ref{tab:TEXONO_detectors} together with a Fano factor of $F_f=0.1096$~\cite{TEXONO:2026eyr}. Following the same procedure as for the COHERENT Ge-mini analysis, the total predicted \cevns and \eves event spectra are obtained by summing the contributions from all naturally occurring germanium isotopes according to their natural abundances and combining the D70 and D50 detector responses. The reconstructed spectrum is then evaluated in 20 uniformly spaced electron equivalent energy bins covering the interval $200$--$400~\mathrm{eV_{ee}}$, following the TEXONO data release.

\begin{figure}[ht!]
    \centering
    \includegraphics[width=0.75\linewidth]{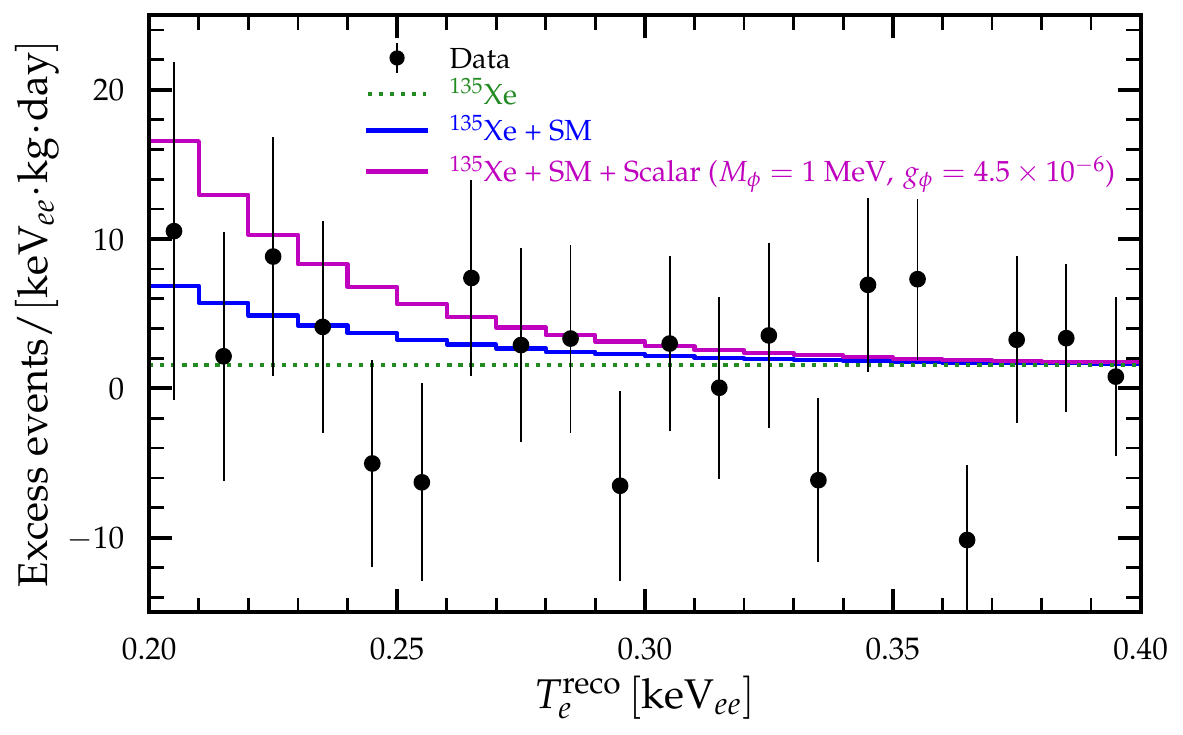}
    \caption{\footnotesize Reactor ON$-$OFF spectrum of the combined D70 and D50 datasets of the TEXONO experiment. The black points represent the measured excess events~\cite{TEXONO:2026eyr}. The green histogram shows the predicted reactor induced $^{135}$Xe Compton background~\cite{TEXONO:2026eyr}, while the blue (magenta) histogram corresponds to the predicted total excess spectrum including the $^{135}$Xe background together with the SM (SM+scalar) contribution. The scalar benchmark parameters are taken to be $M_\phi=1~\mathrm{MeV}$ and $\textsl{g}_\phi=4.5\times10^{-6}$. The predicted spectra include contributions from both \cevns and E\textnu ES.}
    \label{fig:TEXONO_spectrum}
\end{figure}

Figure~\ref{fig:TEXONO_spectrum} shows the reactor ON$-$OFF spectrum of the TEXONO experiment as a function of the reconstructed electron equivalent recoil energy, $T_e^{\rm reco}$. The green histogram represents the predicted reactor induced $^{135}$Xe Compton background, while the black data points correspond to the measured excess events~\cite{TEXONO:2026eyr}. The blue and magenta histograms denote the predicted total excess spectra, including contributions from the $^{135}$Xe background together with the SM and a representative scalar mediated new physics scenario with benchmark parameters $M_\phi=1~\mathrm{MeV}$ and $\textsl{g}_\phi=4.5\times10^{-6}$, respectively.

To constrain the model parameters, we employ the following Gaussian $\chi^2$ test statistic~\cite{TEXONO:2026eyr}:
\begin{equation}
\label{eq:chi2_TEXONO}
\chi^2(\boldsymbol{\Theta})
=
\sum_{i=1}^{20}
\left(
\frac{
R_i^{\rm th}(\boldsymbol{\Theta})
-
R_i^{\rm exp}}
{\sigma_i^{\rm exp}}
\right)^2
+
\left(
\frac{\beta-R^{^{135}\mathrm{Xe}}}
{\sigma^{^{135}\mathrm{Xe}}}
\right)^2,
\end{equation}
where $R_i^{\rm exp}$ and $\sigma_i^{\rm exp}$ denote the measured excess events and their corresponding statistical uncertainties in the $i^\mathrm{th}$ energy bin reported by the TEXONO collaboration~\cite{TEXONO:2026eyr}, and $\boldsymbol{\Theta}$ denotes the set of model parameters under consideration. The predicted number of excess events is given by
\begin{equation}
\label{eq:Rth_TEXONO}
R_i^{\rm th}(\boldsymbol{\Theta})
=
R_i^{\nu\mathcal N}(\boldsymbol{\Theta})
+
R_i^{\nu e}(\boldsymbol{\Theta})
+
\beta,
\end{equation}
where $R_i^{\nu\mathcal N}$ and $R_i^{\nu e}$ denote the predicted \cevns and \eves event rates, respectively, calculated using the formalism stated above. The parameter $\beta$ represents the normalization of the reactor induced $^{135}$Xe Compton background. Following the TEXONO analysis~\cite{TEXONO:2026eyr}, we constrain this background with a Gaussian prior centered at $R^{^{135}\mathrm{Xe}}=1.55$ and a standard deviation of $\sigma^{^{135}\mathrm{Xe}}=0.02$, obtained from a fit to the combined D50 and D70 datasets. For every point in the parameter space, $\boldsymbol{\Theta}$, we minimize Eq.~\eqref{eq:chi2_TEXONO} with respect to the nuisance parameter $\beta$. The minimized $\chi^2$ is then used to determine the corresponding confidence limits.

\section{\label{Sec:Results}Results}

In this section, we present the constraints on the SM and BSM scenarios discussed in Sec.~\ref{Sec:Theory} using the COHERENT Ge-mini and TEXONO datasets.\\[0.1cm]

\begin{figure}[ht!]
    \centering
    \includegraphics[width=0.5\linewidth]{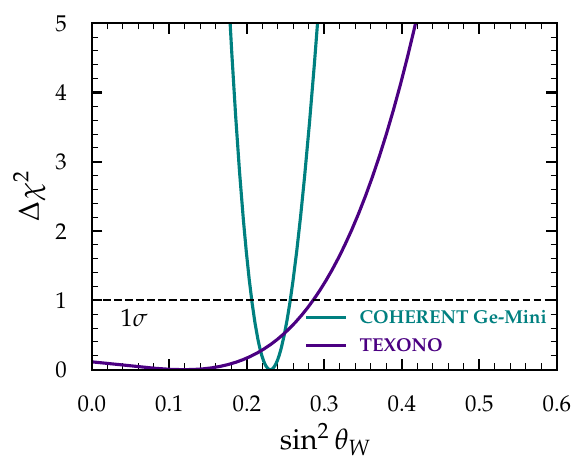}
    \caption{\footnotesize $\Delta\chi^2$ profiles of the weak mixing angle obtained from the COHERENT Ge-mini and TEXONO analyses. The horizontal dashed line indicates the $1\sigma$ confidence level.}
    \label{fig:Delta_chi2_profile_sW2}
\end{figure}

\noindent\textbf{Weak mixing angle:}  We begin with the determination of the weak mixing angle, $\sin^2\theta_W$, one of the fundamental parameters of the electroweak theory. Figure~\ref{fig:Delta_chi2_profile_sW2} shows the corresponding $\Delta\chi^2$ profiles obtained from the two experiments. From these profiles, we obtain the following $1\sigma$ constraints:
\begin{equation*}
\sin^2\theta_W
\left\{
\begin{aligned}
&= {0.231^{+0.025}_{-0.024}}
&& \text{(COHERENT Ge-mini)}\,,\\
&\le 0.285
&& \text{(TEXONO)}\,.
\end{aligned}
\right.
\end{equation*}
The COHERENT Ge-mini dataset yields a two-sided determination of the weak mixing angle centered at $\sin^2\theta_W={0.231}$, which is consistent with the SM prediction at low momentum transfer. In contrast, the present TEXONO data do not exhibit sufficient sensitivity to establish a lower bound and therefore provide only an upper limit at the $1\sigma$ confidence level. Finally, we note that our determination of the weak mixing angle from the COHERENT Ge-mini dataset is in good agreement with the recent measurements reported in Refs.~\cite{AtzoriCorona:2026wbu,DeRomeri:2026dac}. Furthermore, to the best of our knowledge, this work presents the first constraint on the weak mixing angle using recently reported TEXONO 2026 \cevns data.

\begin{figure}[ht!]
    \centering
    \includegraphics[width=0.75\linewidth]{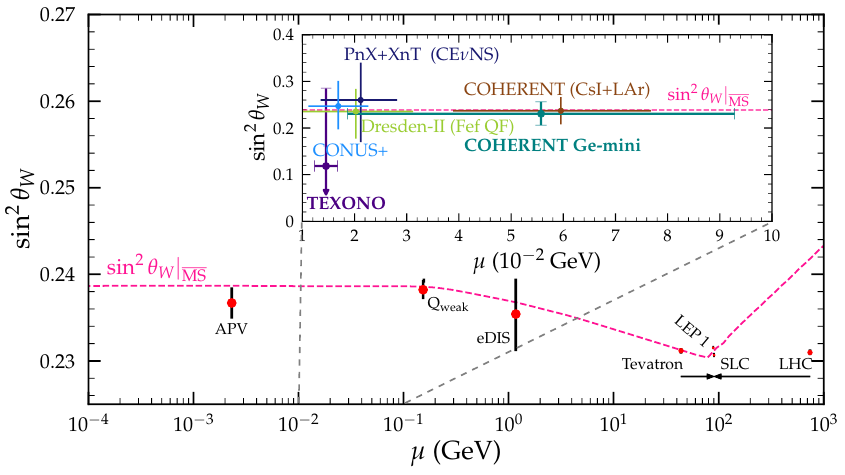}
    \caption{\footnotesize Running of the weak mixing angle, $\sin^2\theta_W$, in the Standard Model under the $\overline{\mathrm{MS}}$ renormalization scheme (magenta dashed line) as a function of the renormalization scale, $\mu$. The teal (purple) error bar denotes the $1\sigma$ determination of $\sin^2\theta_W$ obtained from the COHERENT Ge-mini (TEXONO) data. Other existing experimental measurements spanning a broad range of renormalization scales are included for comparison. Notably, for visual clarity, the Tevatron and LHC points are slightly shifted horizontally with respect to the $Z$ pole mass scale.}
    \label{fig:sW2_Comparison}
\end{figure}

Figure~\ref{fig:sW2_Comparison} compares our determinations of the weak mixing angle with existing measurements spanning a wide range of momentum transfers~\cite{Majumdar:2022nby,DeRomeri:2022twg,DeRomeri:2024iaw,ParticleDataGroup:2024cfk,Chattaraj:2025fvx}. The running of the weak mixing angle predicted in the $\overline{\rm MS}$ scheme is also shown for reference. The precision achieved by the COHERENT Ge-mini measurement is comparable to that of the combined COHERENT CsI+LAr analysis~\cite{DeRomeri:2022twg}. Furthermore, it improves upon the precision of previous reactor based \cevns measurements, including CONUS+~\cite{Chattaraj:2025fvx} and Dresden-II~\cite{Majumdar:2022nby}, as well as the recent solar $^8$B neutrino induced \cevns measurements by PandaX-4T and XENONnT~\cite{DeRomeri:2024iaw}. Although the present TEXONO result is less restrictive than the existing \cevns determinations, it provides an independent measurement using the recently reported TEXONO \cevns data.

\begin{figure}[ht!]
    \centering
        \includegraphics[width=0.49\textwidth]{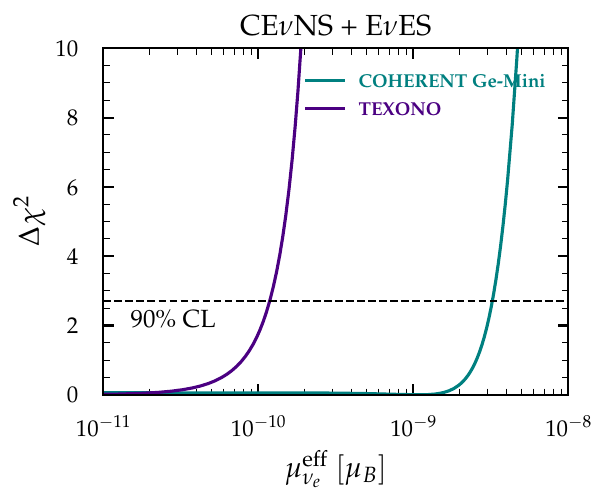}
        \includegraphics[width=0.49\textwidth]{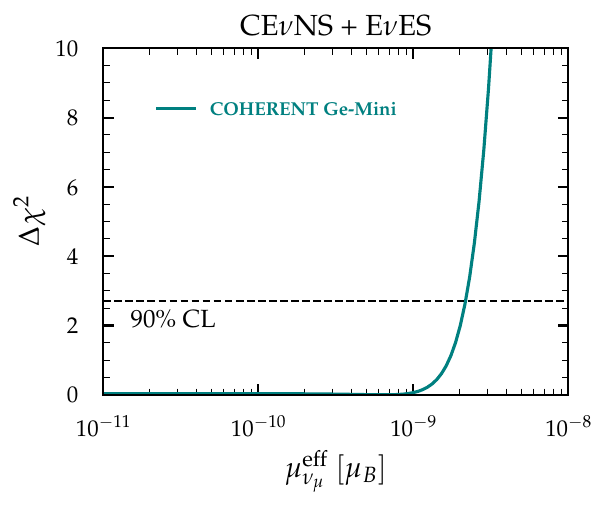}
    \caption{\footnotesize $\Delta\chi^2$ profiles of the effective neutrino magnetic moments. Here, left panel shows results for $\mu_{\nu_e}^{\rm eff}$   obtained from the COHERENT Ge-mini (teal) and TEXONO (purple) data, whereas right panel is for $\mu_{\nu_\mu}^{\rm eff}$ obtained only for the COHERENT Ge-mini (teal) using the combined \cevns and \eves signals.  The horizontal dashed line indicates the $90\%$ CL.}
    \label{Fig:vMM}
\end{figure}

\bigskip

\noindent\textbf{Neutrino magnetic moment:}  We next investigate the constraints on the effective neutrino magnetic moments using the COHERENT Ge-mini and TEXONO datasets, exploiting both \cevns and \eves channels. Since the TEXONO experiment employs reactor antineutrinos, it is sensitive only to the effective electron neutrino magnetic moment, $\mu_{\nu_e}^{\rm eff}$. In contrast, the $\pi$-DAR and $\mu$-DAR neutrino flux at COHERENT Ge-mini contains both electron and muon flavor neutrinos, allowing to put constraints on both $\mu_{\nu_e}^{\rm eff}$ and $\mu_{\nu_\mu}^{\rm eff}$. The corresponding $\Delta\chi^2$ profiles are presented in Figure~\ref{Fig:vMM}. Combining the \cevns and \eves channels, we obtain the following $90\%$ confidence level (CL) upper limits:
\begin{equation*}
\begin{aligned}
\mu_{\nu_e}^\mathrm{eff}
&\leq
\left\{
\begin{aligned}
&{32.35}\times10^{-10}~\mu_B
&& \text{(COHERENT Ge-mini)}\,,\\
&{1.19}\times10^{-10}~\mu_B
&& \text{(TEXONO)}\,,
\end{aligned}
\right.\\[1.5ex]
\mu_{\nu_\mu}^\mathrm{eff}
&\leq
{21.73}\times10^{-10}~\mu_B
\qquad
\text{(COHERENT Ge-mini)}\,.
\end{aligned}
\end{equation*}
For comparison, using only the \cevns signal yields the following $90\%$ CL limits:
\begin{equation*}
\begin{aligned}
\mu_{\nu_e}^\mathrm{eff}
&\leq
\left\{
\begin{aligned}
&{32.70}\times10^{-10}~\mu_B
&& \text{(COHERENT Ge-mini)}\,,\\
&{5.96}\times10^{-10}~\mu_B
&& \text{(TEXONO)}\,,
\end{aligned}
\right.\\[1.5ex]
\mu_{\nu_\mu}^\mathrm{eff}
&\leq
{21.91}\times10^{-10}~\mu_B
\qquad
\text{(COHERENT Ge-mini)}\,.
\end{aligned}
\end{equation*}
The inclusion of the \eves contribution leads to a modest improvement in the sensitivity to the effective neutrino magnetic moments. This improvement is particularly pronounced for TEXONO, where the relatively low energy reactor antineutrino spectrum enhances the sensitivity of \eves to the neutrino magnetic moment. The first column of Table~\ref{tab:EM_limits} compares our $90\%$ CL limits on $\mu_{\nu_e}^{\rm eff}$ and $\mu_{\nu_\mu}^{\rm eff}$ with those reported by other experiments. The TEXONO bound is stronger than the existing \cevns limits from COHERENT (CsI+LAr) and Dresden-II, and is comparable to the recent CONUS+ result. Nevertheless, it remains less stringent than the limits derived from dedicated \eves measurements of solar neutrinos, such as those from XENONnT, LZ, and Borexino. Likewise, although the COHERENT Ge-mini data improve upon the previous constraints obtained from the combined COHERENT CsI and LAr datasets, their sensitivity is still significantly weaker than that achieved by reactor based \cevns measurements and dedicated \eves experiments. Finally, we note that the CE\textnu NS only limit on $\mu_{\nu_e}^{\rm eff}$ obtained in this work using the TEXONO dataset is consistent with the result reported in Ref.~\cite{TEXONO:2026eyr}.

\begin{table*}[ht!]
\centering
\renewcommand{\arraystretch}{1.15}

\begin{adjustbox}{max width=\textwidth}
\setlength{\tabcolsep}{5pt}
\begin{tabular}{cccc}
\toprule
\toprule
\textbf{Flavor} &
{\boldmath$|\mu_\nu^{\mathrm{eff}}|~(10^{-11}\mu_B)$} &
{\boldmath$q_\nu~(10^{-12}e)$} &
{\boldmath$\langle r_\nu^2\rangle~(10^{-32}\,\mathrm{cm}^2)$} \\
\midrule

\multirow{9}{*}{$\nu_e$}
& $\bm{\leq{323.5}}$ \textbf{(COHERENT Ge-mini)}
& {$\bm{[-709,717]}$} \textbf{(COHERENT Ge-mini)}
& ${\bm{[-60.17,-30.65]\cup[-22.27,7.20]}}$ \textbf{(COHERENT Ge-mini)} \\
& \textbf{$\bm{\leq{11.9}}$ (TEXONO)}
& $\bm{[-1.94,2.04]}$ \textbf{(TEXONO)}
& {$\bm{[-62.18, 9.37]}$} \textbf{(TEXONO)} \\
& $\leq1.4$ (LZ)~\cite{A:2022acy}
& $[-0.3,0.6]$ (LZ)~\cite{A:2022acy}
& $[-121,37.5]$ (LZ)~\cite{A:2022acy} \\
& $\leq0.9$ (XENONnT)~\cite{A:2022acy}
& $[-0.1,0.6]$ (XENONnT)~\cite{A:2022acy}
& $[-93.4,9.5]$ (XENONnT)~\cite{A:2022acy} \\
& $\leq420$ (COHERENT CsI+LAr)~\cite{AtzoriCorona:2022qrf}
& $[-500,500]$ (COHERENT CsI+LAr)~\cite{AtzoriCorona:2022qrf}
& $[-69.3,-49.2]\cup [-6.9,14.4]$ (COHERENT CsI+LAr)~\cite{AtzoriCorona:2022qrf} \\
& $\leq11.2$ (CONUS+)~\cite{Chattaraj:2025fvx}
& $[-1.8,1.9]$ (CONUS+)~\cite{Chattaraj:2025fvx}
& $[-59.76,8.33]$ (CONUS+)~\cite{Chattaraj:2025fvx} \\
& $\leq20.8$ (DRESDEN-II)~\cite{AtzoriCorona:2022qrf}
& $[-8.6,8.7]$ (DRESDEN-II)~\cite{AtzoriCorona:2022qrf}
& $[-56.7,-40.8]\cup [-11.6, 4]$ (DRESDEN-II)~\cite{AtzoriCorona:2022qrf} \\
& $\leq2.9$ (GEMMA)~\cite{Beda:2012zz}
&
& $[-5.94,8.28]$ (LSND)~\cite{LSND:2001akn} \\
& $\leq3.9$ (Borexino)~\cite{Borexino:2017fbd, Coloma:2022umy}
&
& \\
\midrule

\multirow{5}{*}{$\nu_\mu$}
& $\bm{\leq{217.3}}$ \textbf{ (COHERENT Ge-mini)}
& {$\bm{[-495,503]}$} \textbf{(COHERENT Ge-mini)}
& {$\bm{[-56.26, -45.88]\cup [-7.05, 3.28]}$} \textbf{(COHERENT Ge-mini)} \\
& $\leq2.3$ (LZ)~\cite{A:2022acy}
& $[-0.7,0.7]$ (LZ)~\cite{A:2022acy}
& $[-109,112.3]$ (LZ)~\cite{A:2022acy} \\
& $\leq1.5$ (XENONnT)~\cite{A:2022acy}
& $[-0.6,0.6]$ (XENONnT)~\cite{A:2022acy}
& $[-50.2,54]$ (XENONnT)~\cite{A:2022acy} \\
& $\leq5.8$ (Borexino)~\cite{Borexino:2017fbd, Coloma:2022umy}
& $\leq11$ (XMASS-I)~\cite{XMASS:2020zke}
& $[-1.2,1.2]$ (CHARM-II)~\cite{CHARM-II:1994aeb} \\
& $\leq 180$ (COHERENT CsI+LAr)~\cite{AtzoriCorona:2022qrf}
& $[-190, 190]$ (COHERENT CsI+LAr)~\cite{AtzoriCorona:2022qrf}
& $[-57.7,-47.8]\cup [-8.8, 3.2]$ (COHERENT CsI+LAr)~\cite{AtzoriCorona:2022qrf} \\
\bottomrule
\bottomrule
\end{tabular}
\end{adjustbox}

\caption{\footnotesize Comparison of the current 90\% CL limits on the flavor diagonal neutrino magnetic moment, millicharge, and charge radius. The first two rows of the $\nu_e$ block are reserved for the new COHERENT Ge-mini and TEXONO results, respectively, while the first row of the $\nu_\mu$ block corresponds to the new COHERENT Ge-mini result.}
\label{tab:EM_limits}
\end{table*}

\begin{figure}[ht!]
    \centering
        \includegraphics[width=0.49\textwidth]{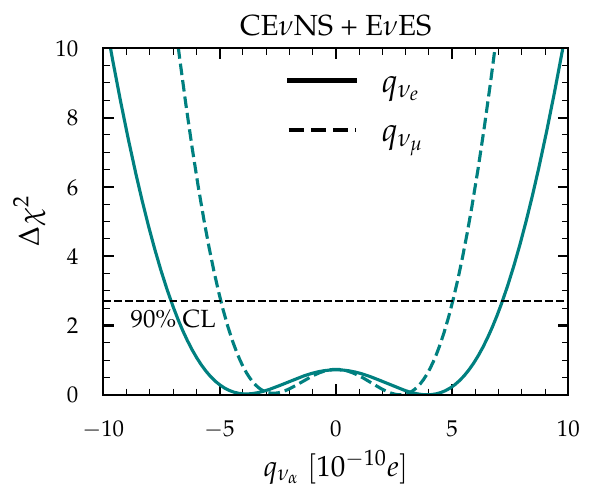}
        \includegraphics[width=0.49\textwidth]{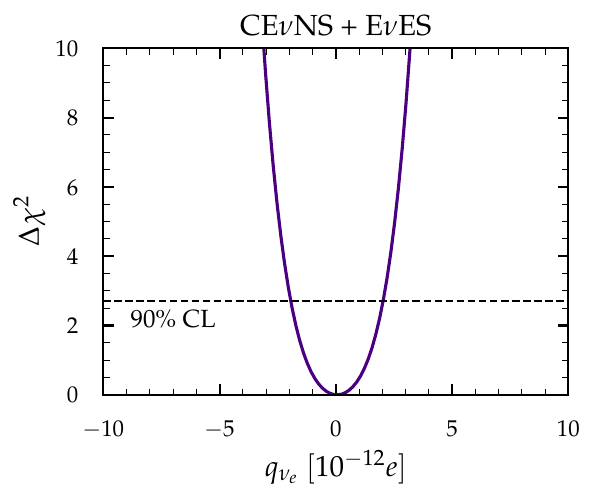}
    \caption{\footnotesize $\Delta\chi^2$ profiles of the neutrino millicharges. Here, the left panel shows results for $q_{\nu_e}$ (solid) and $q_{\nu_\mu}$ (dashed) obtained from the COHERENT Ge-mini data (teal), whereas the right panel is for $q_{\nu_e}$ (solid) obtained from the TEXONO data (purple) using the combined \cevns and \eves signals. The horizontal dashed line indicates the $90\%$ CL.}
    \label{Fig:vQ}
\end{figure}

\bigskip

\noindent\textbf{Neutrino millicharge:}  We next investigate the constraints on the neutrino millicharges using the COHERENT Ge-mini and TEXONO datasets. As in the case of the neutrino magnetic moment, the TEXONO experiment is sensitive only to the electron neutrino millicharge, $q_{\nu_e}$, whereas the $\pi$-DAR and $\mu$-DAR neutrino fluxes at COHERENT Ge-mini allow us to probe both $q_{\nu_e}$ and $q_{\nu_\mu}$. The corresponding $\Delta\chi^2$ profiles are shown in Figure~\ref{Fig:vQ}. Combining the \cevns and \eves signals, we obtain the following $90\%$ CL intervals:
\begin{equation*}
\begin{aligned}
q_{\nu_e}
&\in
\left\{
\begin{aligned}
&{[-7.09,7.17]}\times10^{-10}~e
&&\text{(COHERENT Ge-mini)}\,,\\
&[-1.94,2.04]\times10^{-12}~e
&&\text{(TEXONO)}\,,
\end{aligned}
\right.\\[1.5ex]
q_{\nu_\mu}
&\in
{[-4.95,5.03]}\times10^{-10}~e
\qquad
\text{(COHERENT Ge-mini)}\,.
\end{aligned}
\end{equation*}
For comparison, using only the \cevns signal yields the following $90\%$ CL intervals:
\begin{equation*}
\begin{aligned}
q_{\nu_e}
&\in
\left\{
\begin{aligned}
&{\left[0.97,1.81\right]}\times10^{-7}~e
&& \text{(COHERENT Ge-mini)}\,,\\
&{\left[-6.27,39.36\right]}\times10^{-9}~e
&& \text{(TEXONO)}\,,
\end{aligned}
\right.\\[1.5ex]
q_{\nu_\mu}
&\in
{\left[-0.27,0.20\right]}\times10^{-7}~e
\qquad\qquad\qquad\qquad\qquad\qquad~
\text{(COHERENT Ge-mini)}\,.
\end{aligned}
\end{equation*}
Unlike the case of the neutrino magnetic moment, the inclusion of the \eves signal leads to a substantial improvement in the sensitivity to the neutrino millicharge for both the COHERENT Ge-mini and TEXONO datasets. This pronounced enhancement originates from the strong sensitivity of low energy neutrino electron scattering to the millicharge interaction. The second column of Table~\ref{tab:EM_limits} compares our $90\%$ CL limits with those reported by other experiments. The TEXONO constraint is more stringent than the existing \cevns limits from COHERENT (CsI+LAr) and Dresden-II, and is comparable to the recent CONUS+ result. Nevertheless, it remains less stringent than the limits derived from dedicated \eves measurements of solar neutrinos, such as those from XENONnT and LZ. The COHERENT Ge-mini analysis provides complementary constraints by simultaneously probing both the electron and muon neutrino millicharges, whereas reactor based experiments are sensitive only to $q_{\nu_e}$.

\begin{figure}[ht!]
    \centering
        \includegraphics[width=0.49\textwidth]{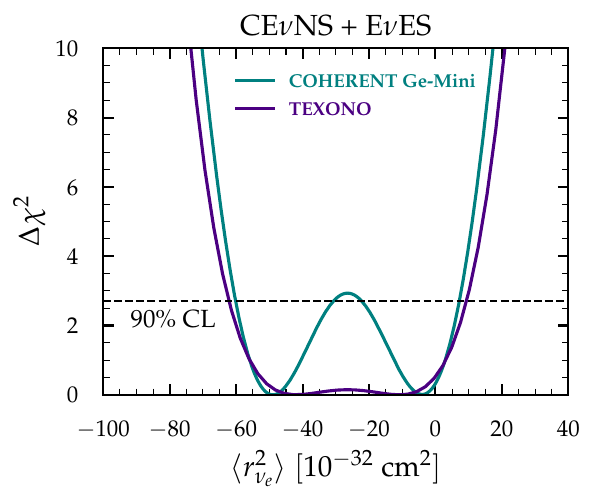}
        \includegraphics[width=0.49\textwidth]{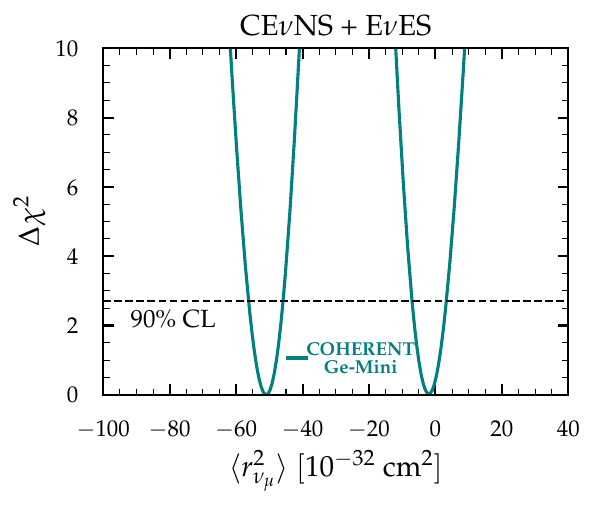}
    \caption{\footnotesize $\Delta\chi^2$ profiles of the neutrino charge radii. Here, the left panel shows results for $\langle r^2_{\nu_e}\rangle$ obtained from the COHERENT Ge-mini (teal) and TEXONO (purple) data, whereas the right panel is for $\langle r^2_{\nu_\mu}\rangle$ obtained only from the COHERENT Ge-mini data (teal) using the combined \cevns and \eves signals. The horizontal dashed line indicates the $90\%$ CL. {The SM values of the neutrino charge radii given in Eq.~\eqref{eq:CR_SM} lie within the allowed ranges in both panels.}}
    \label{Fig:CR}
\end{figure}

\bigskip

\noindent\textbf{Neutrino charge radius:}  Turning to the neutrino charge radius, the TEXONO experiment probes only the electron neutrino charge radius, $\langle r_{\nu_e}^2\rangle$, whereas the $\pi$-DAR and $\mu$-DAR neutrino fluxes at COHERENT Ge-mini enable simultaneous constraints on both $\langle r_{\nu_e}^2\rangle$ and $\langle r_{\nu_\mu}^2\rangle$. The corresponding $\Delta\chi^2$ profiles are presented in Figure~\ref{Fig:CR}. Combining the \cevns and \eves signals, we obtain the following $90\%$ CL intervals:
\begin{equation*}
\begin{aligned}
\langle r^2_{\nu_e}\rangle
&\in
\left\{
\begin{aligned}
&{\left[-60.17,-30.65\right]\times10^{-32}~\mathrm{cm}^2
\cup
\left[-22.27,7.20\right]\times10^{-32}}~\mathrm{cm}^2
&& \text{(COHERENT Ge-mini)}\,,\\
&{\left[-62.18,9.37\right]}\times10^{-32}~\mathrm{cm}^2
&& \text{(TEXONO)}\,,
\end{aligned}
\right.\\[1.5ex]
\langle r^2_{\nu_\mu}\rangle
&\in
{\left[-56.26,-45.88\right]\times10^{-32}~\mathrm{cm}^2
\cup
\left[-7.05,3.28\right]\times10^{-32}}~\mathrm{cm}^2
\qquad~
\text{(COHERENT Ge-mini)}\,.
\end{aligned}
\end{equation*}
{In all cases, the SM values, $\langle r^2_{\nu_e}\rangle_{\rm SM}\simeq-0.83\times10^{-32}~{\rm cm}^2$ and $\langle r^2_{\nu_\mu}\rangle_{\rm SM}\simeq-0.48\times10^{-32}~{\rm cm}^2$, obtained from Eq.~\eqref{eq:CR_SM}, lie within the allowed ranges.}
Since the contribution of \eves to the neutrino charge radius interaction is negligible, the inclusion of the \eves signal produces effectively no change in the allowed parameter space, yielding limits that are nearly identical to those obtained from a CE\textnu NS only analysis. The third column of Table~\ref{tab:EM_limits} compares our $90\%$ CL limits with those reported by other experiments. The COHERENT Ge-mini constraints on the neutrino charge radii are comparable to the previous results obtained from the combined COHERENT CsI and LAr datasets, as well as recent reactor based \cevns experiments such as CONUS+ and Dresden-II. Moreover, the COHERENT Ge-mini constraints are more stringent than those obtained from dedicated \eves analyses using the solar neutrino data from LZ and XENONnT. On the other hand, the TEXONO analysis provides an independent constraint on $\langle r_{\nu_e}^2\rangle$, that is likewise comparable to other recent reactor based \cevns measurements. Finally, we note that our one dimensional constraints on the neutrino charge radii obtained from the COHERENT Ge-mini dataset are consistent with the corresponding two dimensional allowed regions reported in Ref.~\cite{AtzoriCorona:2026wbu}.\\[0.1cm]

\noindent\textbf{Neutrino anapole moment:}  Before concluding the discussions on the limits on neutrino EM properties, we stress that, in elastic neutrino scattering, the effects of the neutrino charge radius and the anapole moment are phenomenologically indistinguishable, since their contributions to the scattering cross section are related by
$
a_{\nu_\alpha}
=
-\langle r_{\nu_\alpha}^2\rangle/6
$,
as follows from Eq.~\eqref{equn:v_millicharge_CR_charge} [see also Ref.~\cite{Giunti:2014ixa}]. Consequently, the corresponding constraints on the neutrino anapole moment can be directly derived from the charge radius limits using the above relation.

\begin{figure}[ht!]
    \centering
    \includegraphics[width=0.5\linewidth]{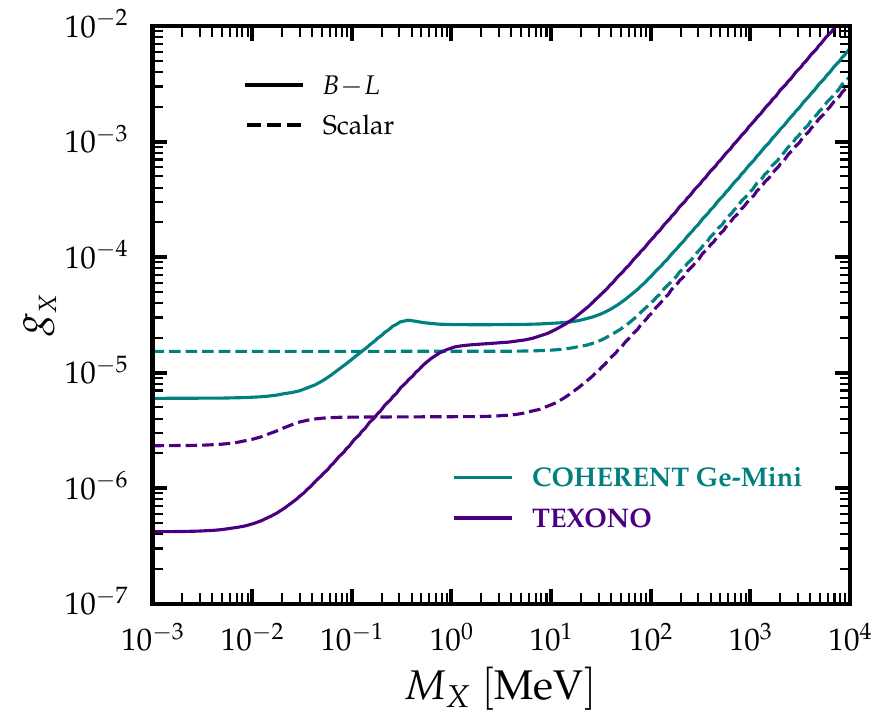}
    \caption{\footnotesize $90\%$ CL upper limits on the vector $B\!-\!L$ gauge coupling, $\textsl{g}_{B\!-\!L}$, and the scalar coupling, $\textsl{g}_{\phi}$, as functions of the mediator masses, $M_{Z^\prime}$ and $M_{\phi}$, respectively, obtained from the COHERENT Ge-mini (teal) and TEXONO (purple) data using the combined \cevns and \eves signals. For the scalar mediator, the benchmark relation $\textsl{g}_{\phi}=\sqrt{g_{\phi}^{\nu}g_{\phi}^{f}}$ with universal fermion-scalar couplings, $g_{\phi}^{u}=g_{\phi}^{d}=g_{\phi}^{e}=g_{\phi}^{f}$, is assumed. Solid (dashed) curves correspond to the vector $B\!-\!L$ (scalar) mediator scenario.}
    \label{fig:NGI}
\end{figure}

\bigskip

\noindent\textbf{Neutrino generalized interactions:}  We now investigate the sensitivities of the COHERENT Ge-mini and TEXONO experiments to light mediators associated with neutrino generalized interactions (NGI). Figure~\ref{fig:NGI} presents the resulting $90\%$ CL upper limits on the vector $B\!-\!L$ gauge coupling, $\textsl{g}_{B\!-\!L}$, and the scalar coupling, $\textsl{g}_{\phi}$, as functions of the corresponding mediator masses, $M_{Z^\prime}$ and $M_{\phi}$, respectively. For the scalar scenario, we adopt the benchmark relation $\textsl{g}_{\phi}=\sqrt{g_{\phi}^{\nu}g_{\phi}^{f}}$, assuming universal fermion-scalar couplings, $g_{\phi}^{u}=g_{\phi}^{d}=g_{\phi}^{e}=g_{\phi}^{f}$. For both the vector $B\!-\!L$ and scalar mediator scenarios, the TEXONO experiment provides stronger constraints than COHERENT Ge-mini in the light mediator regime. This behavior originates from the comparatively lower energy reactor antineutrino spectrum, which enhances the sensitivity to interactions mediated by light particles. In contrast, for the vector $B\!-\!L$ scenario, the COHERENT Ge-mini limits become more stringent at heavier mediator masses owing to the broader neutrino energy spectrum available at the SNS, thereby extending the sensitivity into the heavy mediator regime. A comparison between the two mediator scenarios further reveals that, in the heavy mediator regime, $M_X\gtrsim100~\mathrm{keV}$, the scalar mediator is constrained more strongly than the vector $B\!-\!L$ mediator for both experiments. Finally, the impact of including the \eves signal differs significantly between the two scenarios. In the vector $B\!-\!L$ case, combining \eves with \cevns substantially strengthens the limits from both COHERENT Ge-mini and TEXONO, with the improvement being particularly pronounced for $M_{Z^\prime}\lesssim1~\mathrm{MeV}$. On the other hand, for the scalar mediator, the inclusion of the \eves signal has no visible impact on the COHERENT Ge-mini constraints, while only a slight improvement is observed for TEXONO, primarily in the region $M_{\phi}\lesssim30~\mathrm{keV}$.

\begin{figure}[ht!]
    \centering
        \includegraphics[width=0.49\textwidth]{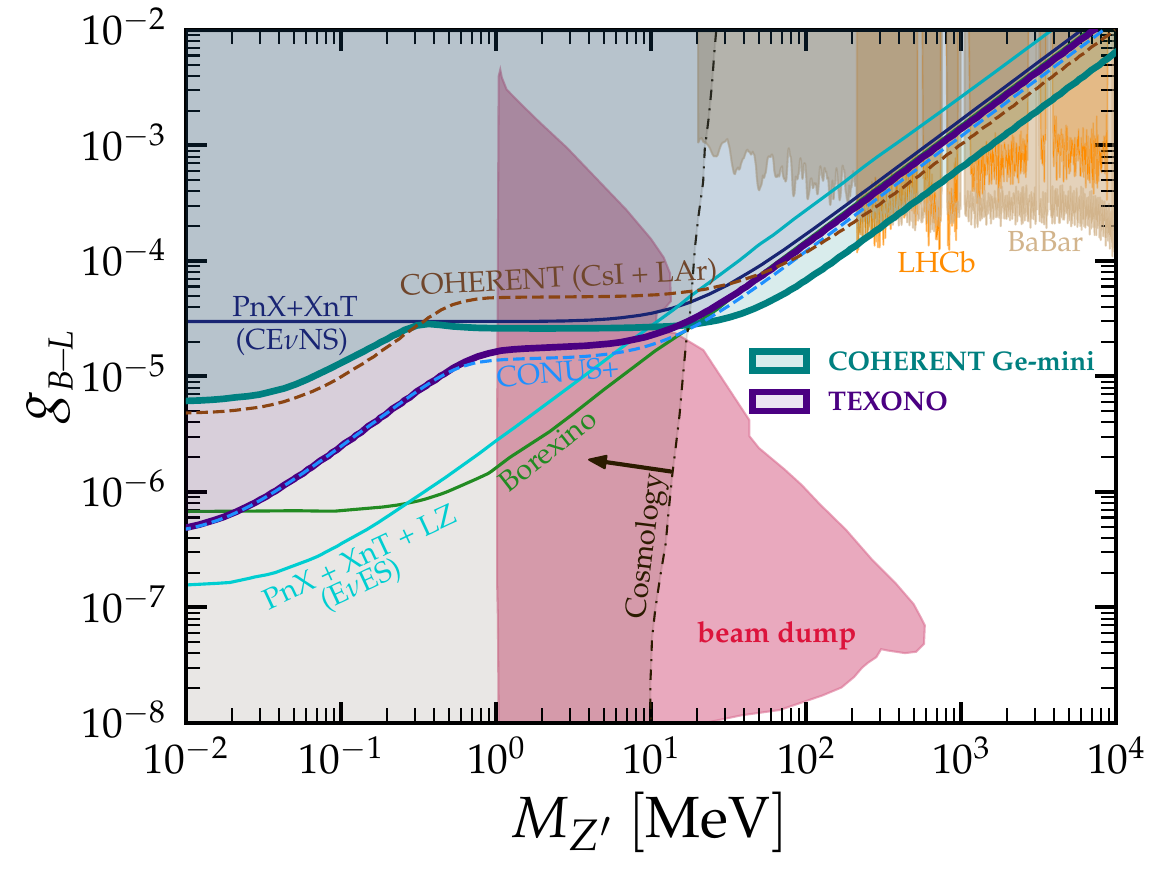}
        \includegraphics[width=0.49\textwidth]{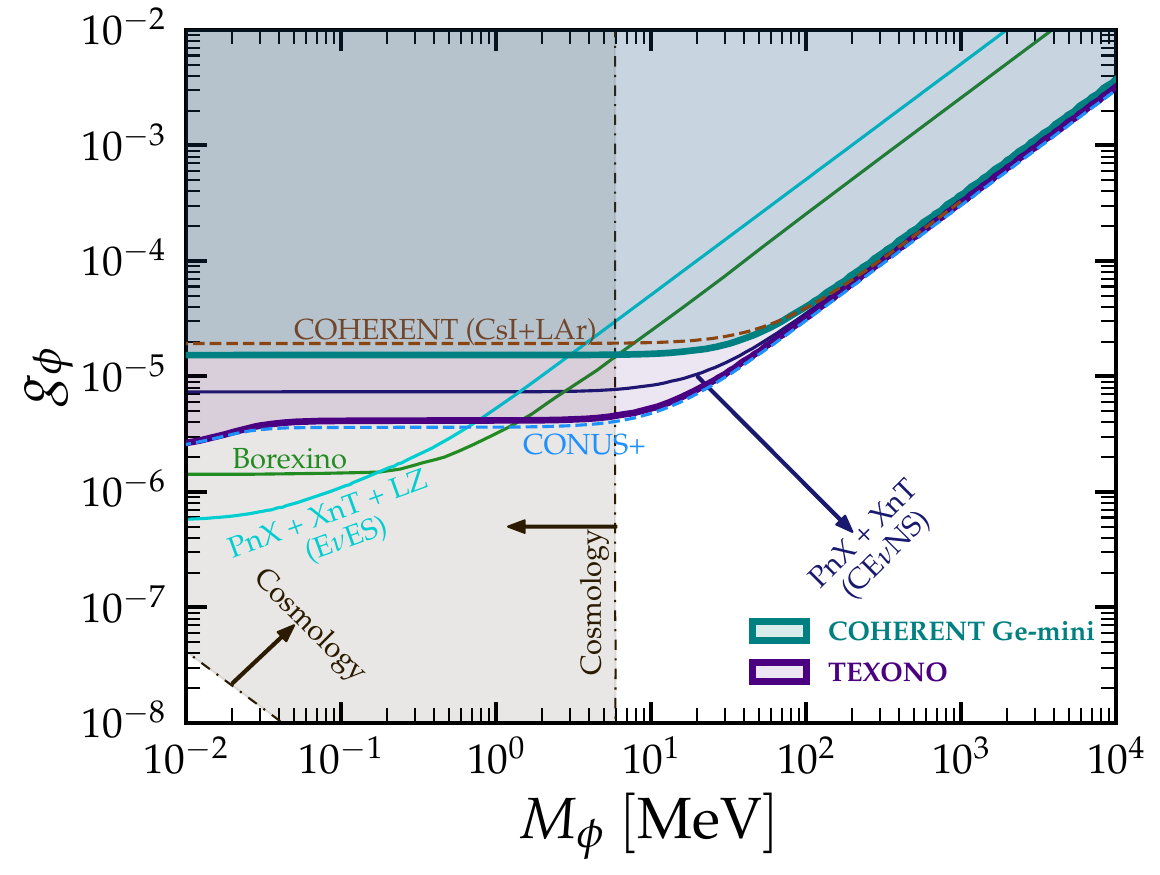}
    \caption{\footnotesize $90\%$ CL upper limits on light mediator couplings. Here, the left panel shows the limits on the vector $B\!-\!L$ gauge coupling, $\textsl{g}_{B\!-\!L}$, in the $(M_{Z^\prime},\textsl{g}_{B\!-\!L})$ plane, whereas the right panel is for the scalar coupling, $\textsl{g}_{\phi}$, in the $(M_{\phi},\textsl{g}_{\phi})$ plane. The COHERENT Ge-mini and TEXONO limits obtained in this work are highlighted by the teal and purple shaded regions, respectively. Existing laboratory, astrophysical, and cosmological constraints are superimposed for comparison.}
    \label{Fig:Light_Mediators}
\end{figure}

A comparison of these results with constraints from other experimental and astrophysical studies further illustrates their complementarity. Figure~\ref{Fig:Light_Mediators} compares the limits obtained in this work with existing constraints in the vector $B\!-\!L$ and scalar mediator parameter spaces. We include exclusion limits derived from the combined analysis of the COHERENT CsI and LAr \cevns data~\cite{DeRomeri:2022twg,AtzoriCorona:2022moj,Majumdar:2024dms}. Notably, the CsI analysis incorporates both \cevns and \eves contributions, whereas the LAr results rely solely on \cevns events. We also show the constraints from the recent CONUS+ analysis~\cite{Chattaraj:2025fvx,DeRomeri:2025csu,AtzoriCorona:2025ygn}, together with the limits derived from recent measurements of solar $^8\mathrm{B}$ neutrino induced \cevns signals by PandaX-4T and XENONnT~\cite{DeRomeri:2024iaw,Blanco-Mas:2024ale}. For completeness, constraints from the \eves channel are also included, namely, those from BOREXINO~\cite{Coloma:2022umy}, as well as the combined analyses of solar \eves data from PandaX-4T, XENONnT, and LZ~\cite{A:2022acy,DeRomeri:2024dbv,Majumdar:2024dms}.
We further showcase limits from dark photon searches at fixed target and beam dump facilities, as well as at high energy colliders such as BaBar and LHCb (see~\cite{Ilten:2018crw} and references therein). These limits have been recast into the corresponding parameter space using the \texttt{darkcast} software package~\cite{darkcast}, following Refs.~\cite{Ilten:2018crw,Baruch:2022esd}.
 Finally, we include astrophysical and cosmological bounds arising from the effective number of relativistic degrees of freedom, $N_{\rm eff}$~\cite{Esseili:2023ldf,Li:2023puz,Ghosh:2024cxi}, together with the Big Bang Nucleosynthesis (BBN) constraints~\cite{Blinov:2019gcj,Suliga:2020jfa}.

The complementarity of the COHERENT Ge-mini and TEXONO sensitivities with other terrestrial, astrophysical, and cosmological probes is particularly evident in Figure~\ref{Fig:Light_Mediators}. Relative to previous COHERENT analyses, the COHERENT Ge-mini results improve the existing COHERENT CsI+LAr constraints for $M_{Z^\prime}\gtrsim200~\mathrm{keV}$ in the vector $B\!-\!L$ scenario, while providing stronger limits throughout the entire scalar mediator parameter space. Furthermore, the TEXONO limits are among the most stringent constraints for the scalar mediator for $M_{\phi}\gtrsim6~\mathrm{MeV}$, together with those obtained from the recent CONUS+ analysis. In the vector $B\!-\!L$ scenario, the COHERENT Ge-mini data yield the strongest constraints in the approximate mass range $10~\mathrm{MeV}\lesssim M_{Z^\prime}\lesssim200~\mathrm{MeV}$.

\begin{figure}[ht!]
    \centering
        \includegraphics[width=0.49\textwidth]{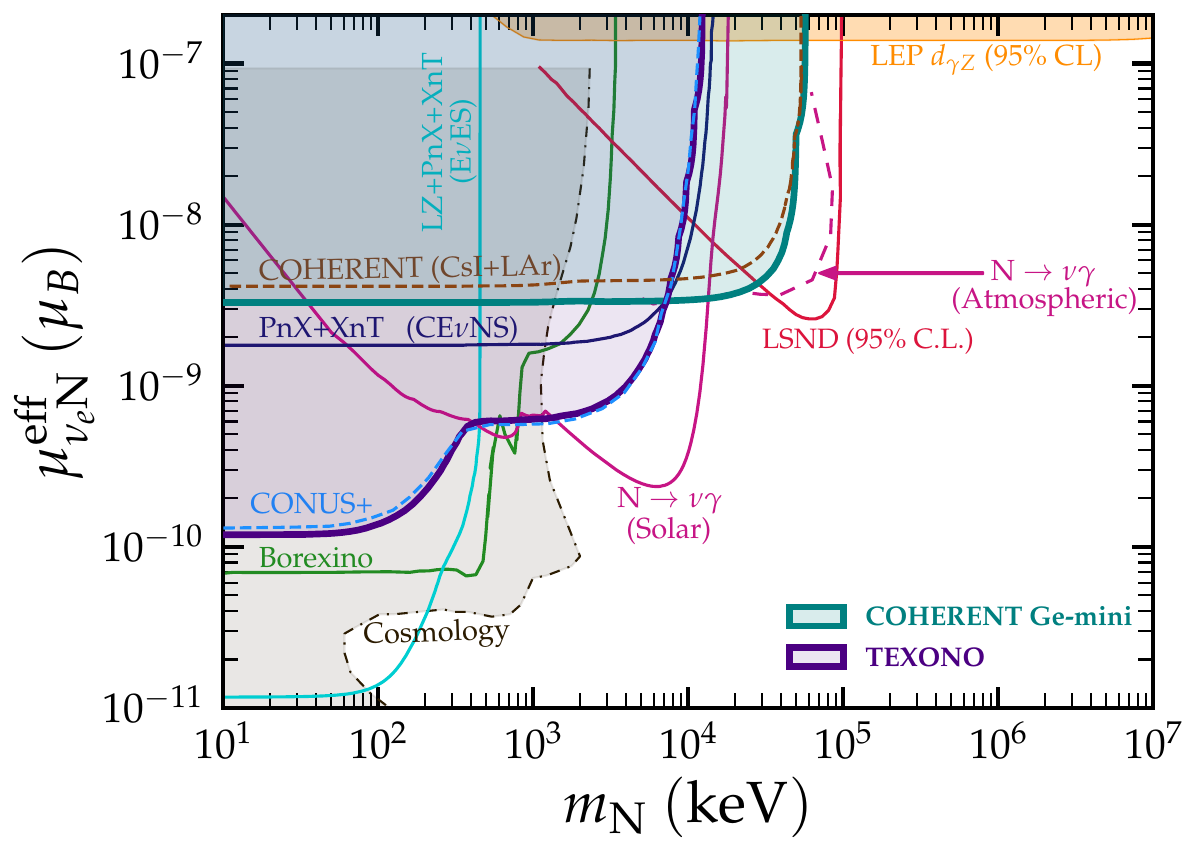}
        \includegraphics[width=0.49\textwidth]{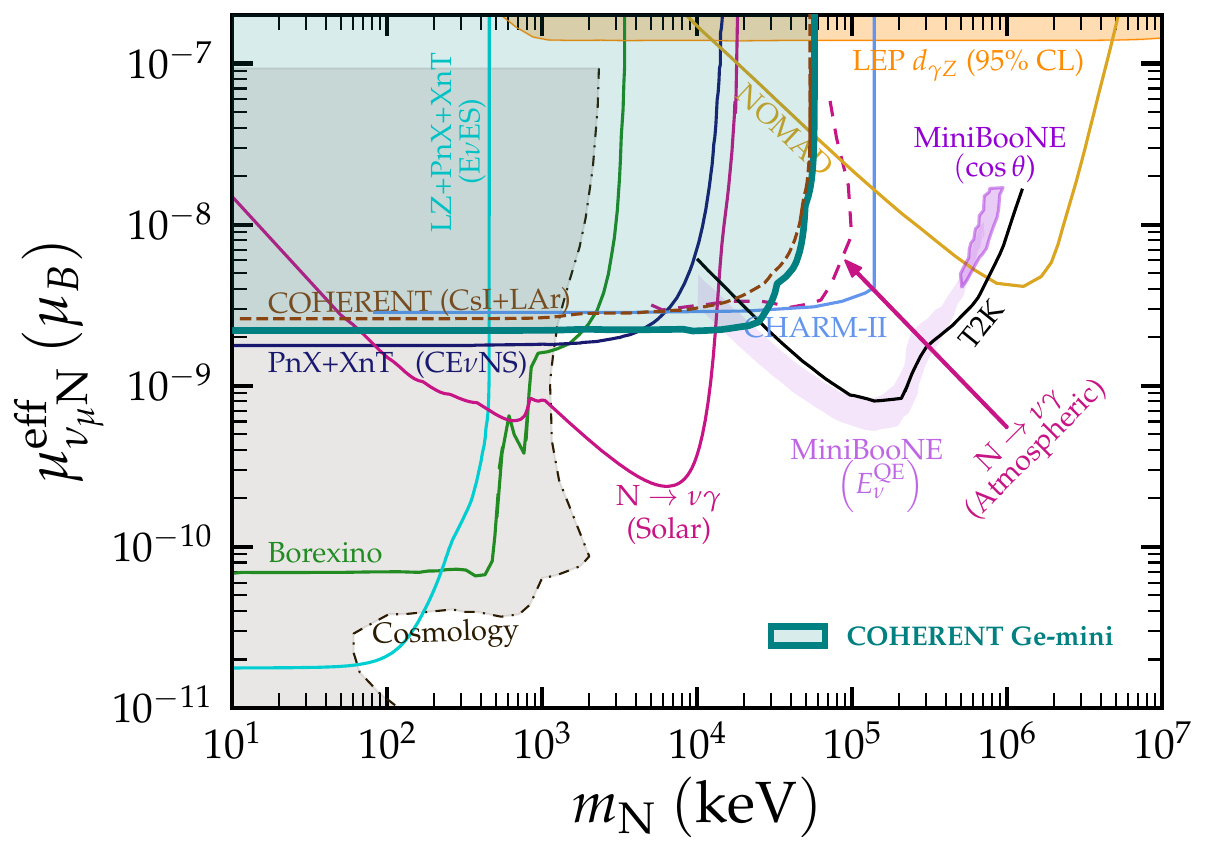}
    \caption{\footnotesize $90\%$ CL exclusion limits on the active-sterile neutrino transition magnetic moments as functions of the SNL mass, $m_{\rm N}$. Here, the left panel shows the limits on $\mu^{\rm eff}_{\nu_e\mathrm{N}}$, whereas the right panel is for $\mu^{\rm eff}_{\nu_\mu\mathrm{N}}$. The COHERENT Ge-mini and TEXONO limits obtained in this work are highlighted by the teal and purple shaded regions, respectively. Existing laboratory, astrophysical, and cosmological constraints are superimposed for comparison.}
    \label{Fig:TMM}
\end{figure}

\bigskip

\noindent\textbf{SNL upscattering via transition dipole portal:}  We next investigate the sensitivity to active-sterile neutrino transitions induced by nonzero transition magnetic moments (TMMs) using the COHERENT Ge-mini and TEXONO data. Figure~\ref{Fig:TMM} presents the resulting $90\%$ CL exclusion limits on $\mu_{\nu_e\mathrm{N}}^{\rm eff}$ (left panel) and $\mu_{\nu_\mu\mathrm{N}}^{\rm eff}$ (right panel) as functions of the SNL mass, $m_{\rm N}$. Owing to the specific recoil energy windows of the COHERENT Ge-mini and TEXONO detectors, together with the energy spectra of the SNS and reactor neutrino fluxes, the kinematically accessible SNL masses are limited to approximately $50$ and $10~\mathrm{MeV}$, respectively, according to the equations discussed in footnotes~\ref{FootNote:Sterile_Mass_Limit_CEvNS} and \ref{FootNote:Sterile_Mass_Limit_EvES}.

For comparison, we include the existing constraints from \cevns analyses of COHERENT (CsI+LAr)~\cite{Miranda:2021kre,DeRomeri:2022twg} and CONUS+~\cite{DeRomeri:2025csu}, together with the combined analysis of PandaX-4T and XENONnT data~\cite{DeRomeri:2024hvc}. We also show limits obtained from \eves analyses using Borexino~\cite{Brdar:2020quo,Plestid:2020vqf} and CHARM-II~\cite{Coloma:2017ppo} data, as well as those from the recent combined analysis of XENONnT, PandaX-4T, and LUX-ZEPLIN data~\cite{DeRomeri:2024hvc}. Additional constraints from LSND~\cite{Magill:2018jla}, LEP~\cite{Magill:2018jla}, NOMAD~\cite{NOMAD:1997pcg, Gninenko:1998nn}, MiniBooNE\footnote{For MiniBooNE, the displayed constraints correspond to muon neutrinos only and are obtained from the combined analysis of reconstructed neutrino energy ($E_\nu^{\rm QE}$) and angular ($\cos\theta$) distributions.}~\cite{Kamp:2022bpt}, and T2K~\cite{T2K:2019jwa, Liu:2024cdi}, and from the radiative decay $\mathrm{N}\rightarrow\nu\gamma$~\cite{Plestid:2020vqf, Plestidlumsolnu}, using solar (Borexino and Super-Kamiokande) and atmospheric (Super-Kamiokande) data~\cite{Gustafson:2022rsz}, are also displayed\footnote{We account for a factor of two difference in the interaction Lagrangian convention adopted in Refs.~\cite{Magill:2018jla,Plestid:2020vqf, Gustafson:2022rsz}.}. Finally, we superimpose the astrophysical and cosmological limits arising from BBN~\cite{Magill:2018jla,Brdar:2020quo} and CMB constraints on $\Delta N_{\rm eff}$~\cite{Brdar:2020quo}.

For SNL masses below $10~\mathrm{MeV}$, the COHERENT Ge-mini analysis performed in this work yields exclusion limits as low as $\mu_{\nu_e\mathrm{N}}^{\rm eff}\sim3\times10^{-9}\,\mu_B$ and $\mu_{\nu_\mu\mathrm{N}}^{\rm eff}\sim2\times10^{-9}\,\mu_B$, while for $m_\mathrm{N}\lesssim 1$ MeV, the corresponding TEXONO analysis reaches $\mu_{\nu_e\mathrm{N}}^{\rm eff}\sim1.2\times10^{-10}\,\mu_B$. The COHERENT Ge-mini limits constitute a modest improvement over the previous constraints derived from the combined COHERENT CsI and LAr data for both the electron and muon neutrino TMMs. Nevertheless, in the low mass region they remain weaker than those obtained from the combined PandaX-4T and XENONnT \cevns analyses, reactor based \cevns experiments such as TEXONO and CONUS+, and dedicated \eves studies based on solar neutrino data, including Borexino, XENONnT, LUX-ZEPLIN, and PandaX-4T. On the other hand, owing to the higher energy SNS neutrino spectrum, the COHERENT Ge-mini experiment extends the accessible SNL mass range well beyond that of reactor experiments. In particular, within the approximate interval $20~\mathrm{MeV}\lesssim m_{\rm N}\lesssim40~\mathrm{MeV}$, the COHERENT Ge-mini analysis provides some of the most stringent constraints on both $\mu_{\nu_e\mathrm{N}}^{\rm eff}$ and $\mu_{\nu_\mu\mathrm{N}}^{\rm eff}$. The limit on $\mu_{\nu_e\mathrm{N}}^{\rm eff}$ obtained from the TEXONO data is comparable to that derived from the recent CONUS+ analysis.

Before concluding, we emphasize that the various limits shown in Figure~\ref{Fig:TMM} are not always directly comparable. The effective transition magnetic moments probed by different experiments generally depends on different combinations of the fundamental TMM couplings, $CP$ violating phases, and neutrino oscillation parameters [see Ref.~\cite{Miranda:2021kre} for a detailed discussion]. A direct comparison is therefore meaningful only among experiments utilizing the same neutrino source, such as COHERENT Ge-mini, COHERENT CsI, and COHERENT LAr, which all exploit the SNS neutrino flux. Likewise, the comparison between the TEXONO and CONUS+ constraints is straightforward, since both experiments employ reactor antineutrinos.

\begin{figure}[ht!]
    \centering
        \includegraphics[width=0.49\textwidth]{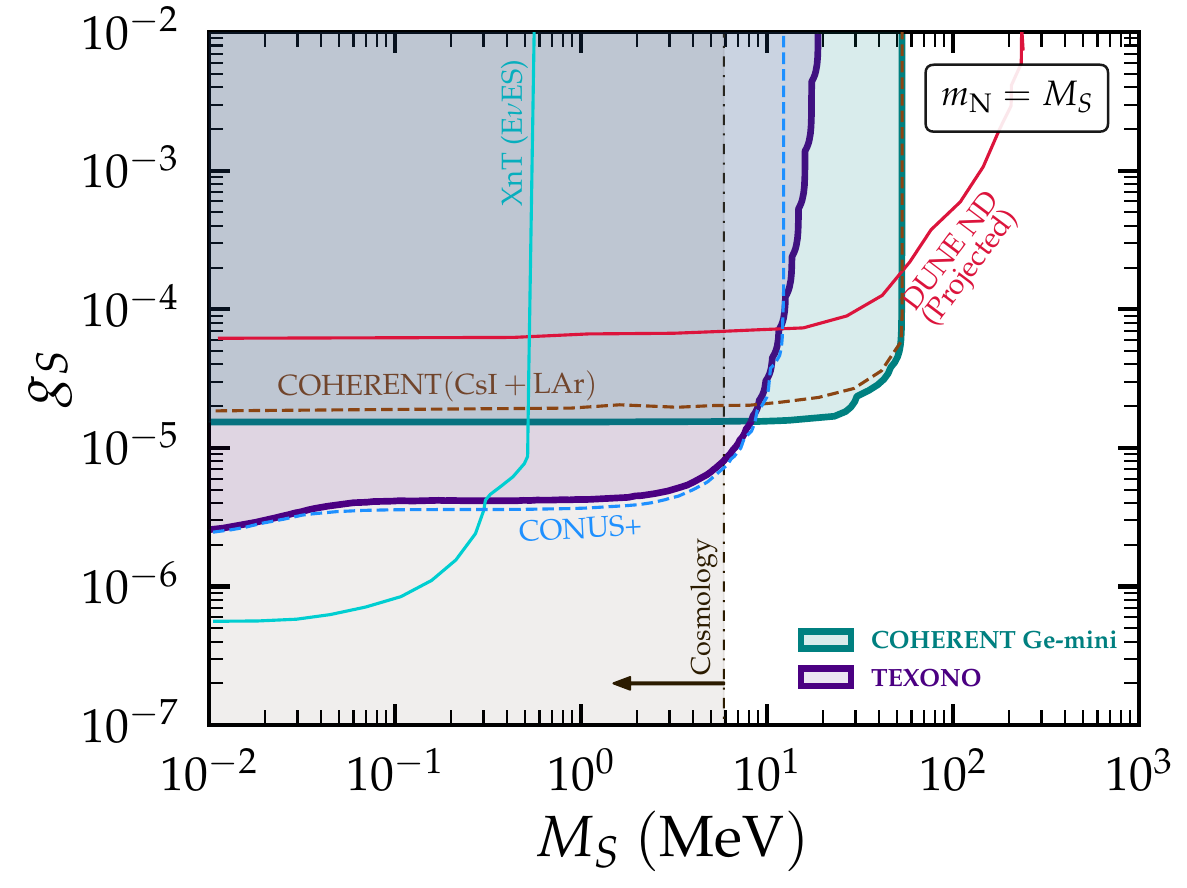}
        \includegraphics[width=0.49\textwidth]{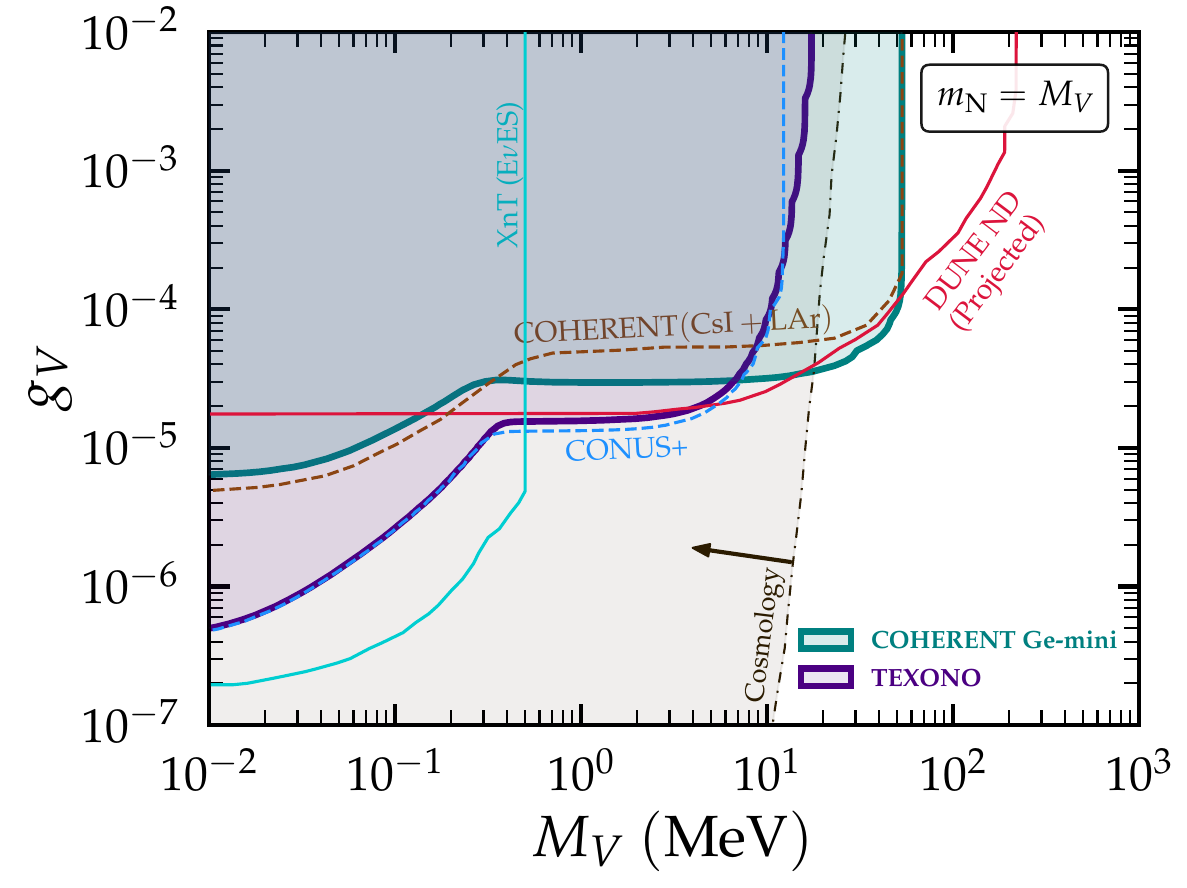}
    \caption{\footnotesize $90\%$ CL exclusion limits on SNL upscattering via generalized interactions for the benchmark $m_{\rm N}=M_X$. Here, the left panel shows the limits for the scalar mediator in the $(M_S,\textsl{g}_S)$ plane, whereas the right panel is for the vector mediator in the $(M_V,\textsl{g}_V)$ plane. The COHERENT Ge-mini and TEXONO limits obtained in this work are highlighted by the teal and purple shaded regions, respectively. Existing laboratory, astrophysical, cosmological, and projected DUNE Near Detector constraints are superimposed for comparison.}
    \label{Fig:Upscattering_SNL}
\end{figure}

\bigskip

\noindent\textbf{SNL upscattering via generalized interactions:}  We now investigate the sensitivity to SNL production via neutrino upscattering mediated by light scalar and vector bosons at the COHERENT Ge-mini and TEXONO experiments. The resulting $90\%$ CL exclusion limits are presented in Figure~\ref{Fig:Upscattering_SNL}. The left and right panels correspond to scalar and vector mediated interactions, respectively. Throughout this figure we adopt the benchmark choice $m_{\rm N}=M_X$, where $X=\{S,V\}$. The corresponding limits for the additional benchmark choices $m_{\rm N}=\{0.1,\,10\}\times M_X$, together with their complementary representations in the $(m_{\rm N},\textsl{g}_X)$ parameter space, are presented in Appendix~\ref{App:SNL_Benchmarks}. In each panel of Figure~\ref{Fig:Upscattering_SNL}, the constraints are shown in the mediator mass-coupling parameter space, $(M_X,\textsl{g}_X)$. In the light mediator regime, the exclusion contours approach a plateau and become nearly identical to the corresponding generalized interaction limits shown in Figure~\ref{Fig:Light_Mediators}. This behavior reflects the fact that, for sufficiently small mediator masses, the kinematic suppression associated with SNL production becomes negligible and the sensitivity is primarily governed by the mediator interaction itself. As the mediator (or equivalently the SNL) mass increases, the sensitivity gradually deteriorates owing to the reduced kinematic phase space available for SNL production. Ultimately, the exclusion contours terminate at the maximum SNL masses allowed by the kinematics discussed in footnotes~\ref{FootNote:Sterile_Mass_Limit_CEvNS} and \ref{FootNote:Sterile_Mass_Limit_EvES}.

To illustrate the complementarity of the obtained sensitivities, Figure~\ref{Fig:Upscattering_SNL} also includes the existing constraints from COHERENT (CsI+LAr) (\cevns and E\textnu ES), XENONnT (E\textnu ES), and the projected sensitivity of the DUNE Near Detector (E\textnu ES), taken from Ref.~\cite{Candela:2024ljb}. The limits derived from the recent CONUS+ analysis~\cite{DeRomeri:2025csu} are also superimposed. Furthermore, whenever available, the corresponding cosmological constraints discussed in Figure~\ref{Fig:Light_Mediators} are also shown.

The comparison clearly demonstrates the complementarity among the various experiments. At very small mediator masses, the XENONnT \eves analysis provides the most stringent limits. In the intermediate mass region, approximately $300~\mathrm{keV}\lesssim m_{\rm N}\lesssim10~\mathrm{MeV}$, the TEXONO limits obtained in this work, together with those from the recent CONUS+ analysis, become more stringent than the XENONnT constraints. Owing to the higher energy SNS neutrino spectrum, the COHERENT Ge-mini experiment substantially extends the accessible SNL mass range beyond that of reactor experiments. In particular, within the approximate interval $10~\mathrm{MeV}\lesssim m_{\rm N}\lesssim50~\mathrm{MeV}$, the COHERENT Ge-mini limits become among the most stringent laboratory constraints and slightly improve upon the previous COHERENT (CsI+LAr) results. For both scalar and vector mediated interactions, the COHERENT Ge-mini and TEXONO sensitivities surpass the projected DUNE Near Detector limits up to their respective kinematic mass reaches. At larger SNL masses, however, the much higher neutrino beam energy available at DUNE considerably extends the accessible parameter space. Consequently, TEXONO, COHERENT Ge-mini, and the future DUNE Near Detector provide highly complementary probes of SNL upscattering, with each experiment offering the leading sensitivity in a different SNL mass regime.

\FloatBarrier

\section{\label{Sec:Conclusions}Conclusions}

\cevns has firmly established itself as a precision tool for testing the SM at low momentum transfer and for probing a broad spectrum of new physics scenarios. In this work, we have performed an extensive statistical analysis of the two most recent \cevns datasets obtained with germanium detectors: the high statistics measurement of the COHERENT Ge-mini detector array at the SNS and the reactor based measurement reported by the TEXONO experiment at KSNL. Our analysis incorporates both the \cevns and \eves channels, a full treatment of the detector responses including quenching, energy resolution, and atomic binding effects, and dedicated Poissonian (COHERENT Ge-mini) and Gaussian (TEXONO) $\chi^2$ test statistics with the relevant systematic uncertainties profiled out.

Within the SM, we have extracted the weak mixing angle at low momentum transfer, obtaining the $1\sigma$ determination $\sin^2\theta_W={0.231^{+0.025}_{-0.024}}$ from the COHERENT Ge-mini data, in excellent agreement with the SM prediction in the low energy regime, with a precision comparable to that of the combined COHERENT CsI+LAr analysis and superior to previous reactor based \cevns determinations. The TEXONO data, while currently providing only an upper limit, $\sin^2\theta_W\le0.285$ at $1\sigma$, yield the first constraint on the weak mixing angle from the recently released TEXONO \cevns data, providing an independent low energy determination of the weak mixing angle.

Turning to BSM physics, we have derived updated constraints on the neutrino electromagnetic properties, namely the effective magnetic moment, millicharge, charge radius, and anapole moment. The TEXONO dataset yields particularly competitive bounds, $\mu_{\nu_e}^{\rm eff}\le{1.19}\times10^{-10}\,\mu_B$ and $q_{\nu_e}\in[-1.94,2.04]\times10^{-12}\,e$ at $90\%$ CL, which improve upon the existing COHERENT CsI+LAr and Dresden-II limits and are comparable to the recent CONUS+ results. The COHERENT Ge-mini data, benefiting from the multi flavor SNS neutrino flux, provides simultaneous constraints on both electron and muon flavor properties, with charge radius limits that are competitive with existing \cevns determinations and more stringent than those derived from solar neutrino \eves data collected by dark matter detectors. We find that the inclusion of the \eves channel is especially consequential for the neutrino millicharge, improving the COHERENT Ge-mini (TEXONO) sensitivity by roughly two (three) orders of magnitude relative to the CE\textnu NS only analysis, a direct consequence of the pronounced $1/\qtransfer^2$ enhancement of the millicharge interaction at low electron recoil energies.

We have further explored neutrino generalized interactions mediated by light vector and scalar bosons, adopting the anomaly free $U(1)_{B\!-\!L}$ gauge extension and a minimal $CP$ even scalar as representative benchmarks. The two experiments exhibit a striking complementarity: TEXONO dominates in the light mediator regime owing to the low energy reactor antineutrino spectrum, whereas COHERENT Ge-mini takes over for heavier mediators thanks to the broader SNS spectrum. In particular, the COHERENT Ge-mini data provide the leading constraints on the vector $B\!-\!L$ coupling in the range $10~\mathrm{MeV}\lesssim M_{Z^\prime}\lesssim200~\mathrm{MeV}$, while the TEXONO limits on the scalar coupling are among the most stringent for $M_\phi\gtrsim6~\mathrm{MeV}$, together with the recent CONUS+ results.

Finally, we have investigated the production of sterile neutral leptons through active to sterile upscattering, considering both the transition dipole portal and generalized scalar and vector interactions. The kinematic reach of the two experiments is again complementary: TEXONO probes SNL masses up to $\sim10~\mathrm{MeV}$ with excellent sensitivity, reaching $\mu_{\nu_e\mathrm{N}}^{\rm eff}\sim1.2\times10^{-10}\,\mu_B$ for $m_{\rm N}\lesssim1~\mathrm{MeV}$, while COHERENT Ge-mini extends the accessible mass range up to $\sim50~\mathrm{MeV}$, delivering some of the most stringent laboratory constraints on the transition magnetic moments in the window $20~\mathrm{MeV}\lesssim m_{\rm N}\lesssim40~\mathrm{MeV}$. For SNL upscattering via generalized interactions, both experiments surpass the projected DUNE Near Detector sensitivity within their respective kinematic reaches, demonstrating that current generation \cevns experiments already probe parameter space beyond the anticipated coverage of forthcoming accelerator based facilities in the low mass regime.

Taken together, our results demonstrate that the combination of stopped pion and reactor based \cevns measurements with germanium detectors constitutes a powerful and mutually complementary probe of both SM precision observables and a wide variety of new physics scenarios, spanning neutrino electromagnetic properties, light mediators, and sterile neutrino portals. With the anticipated accumulation of statistics at the SNS, the deployment of larger detector arrays, and continued progress in lowering thresholds and backgrounds at reactor facilities, the sensitivities reported here are expected to improve significantly in the near future, further consolidating \cevns as a precision laboratory for neutrino physics beyond the SM.

%%%%%%%%%%%%%%%%%%%%%%%%%%%%%%%%%%%
\acknowledgments
%%%%%%%%%%%%%%%%%%%%%%%%%%%%%%%%%%%%
\begin{justify}We are grateful to Dimitrios K. Papoulias for valuable discussions regarding the extraction of the COHERENT Ge-mini data used in this work. A.M. acknowledges financial support from the Government of India through the Prime Minister's Research Fellowship (PMRF), ID: 0401970.\end{justify}

\appendix
\section{Impact of the benchmark relationship between $m_{\rm N}$ and $M_X$ on SNL upscattering via generalized interactions}
\label{App:SNL_Benchmarks}

\begin{figure}[ht!]
    \centering
        \includegraphics[width=0.49\textwidth]{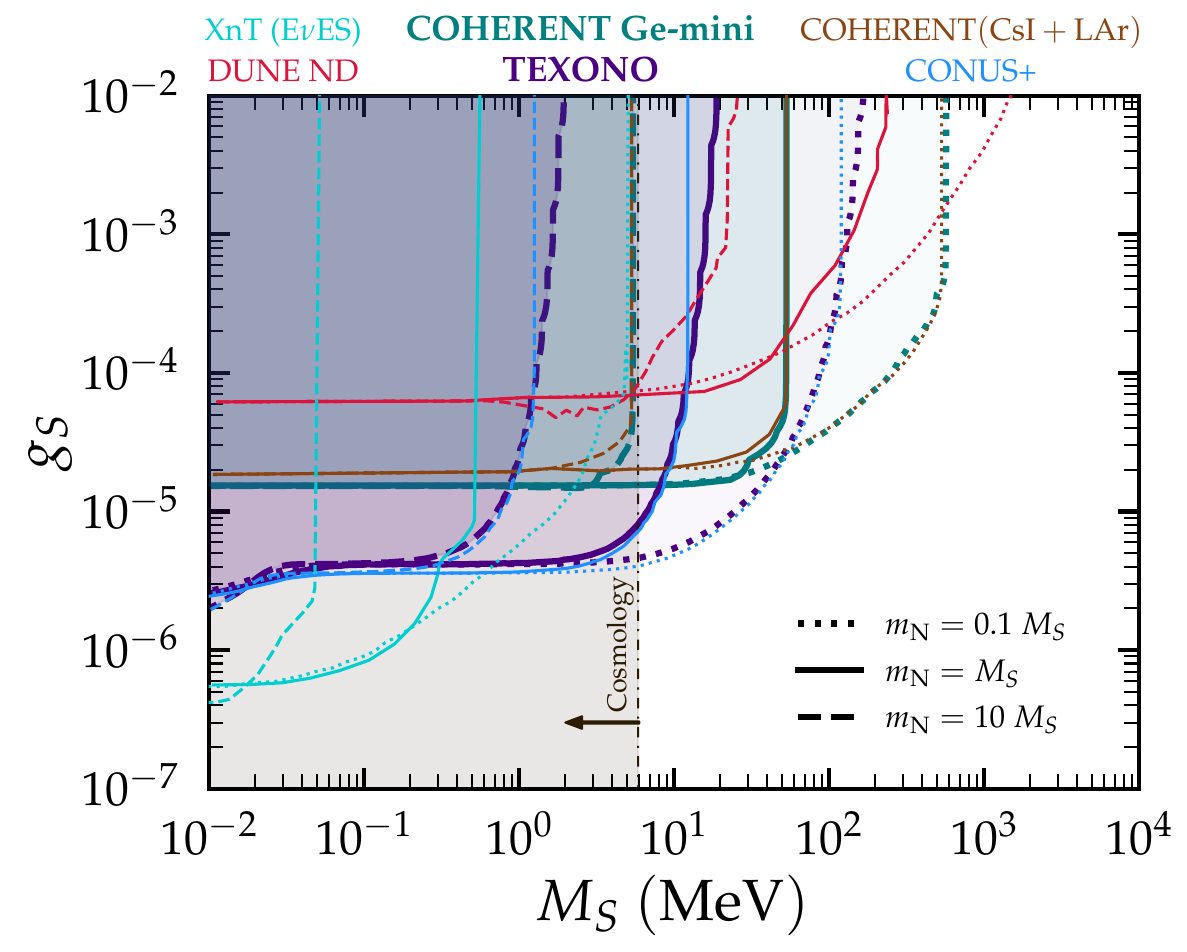}
        \includegraphics[width=0.49\textwidth]{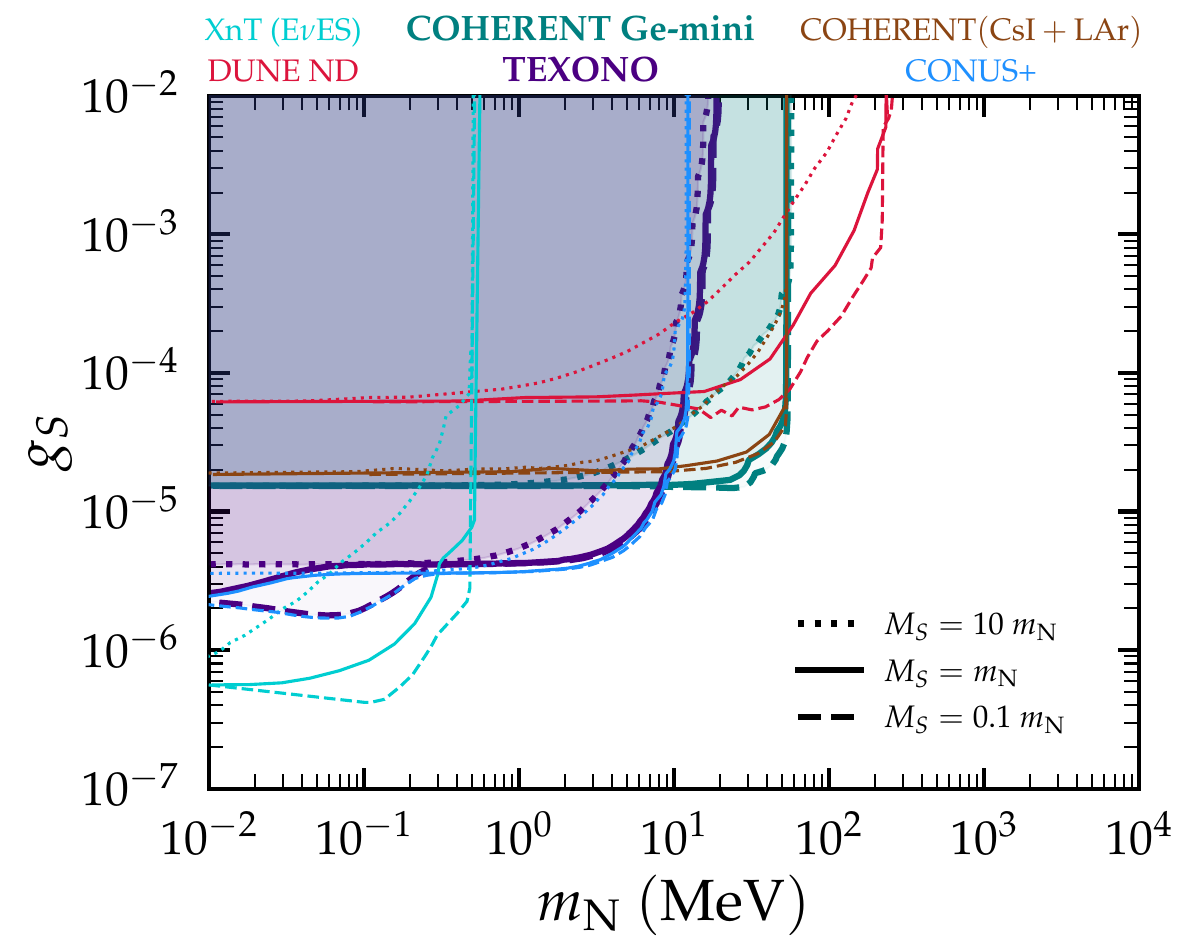}
        \includegraphics[width=0.49\textwidth]{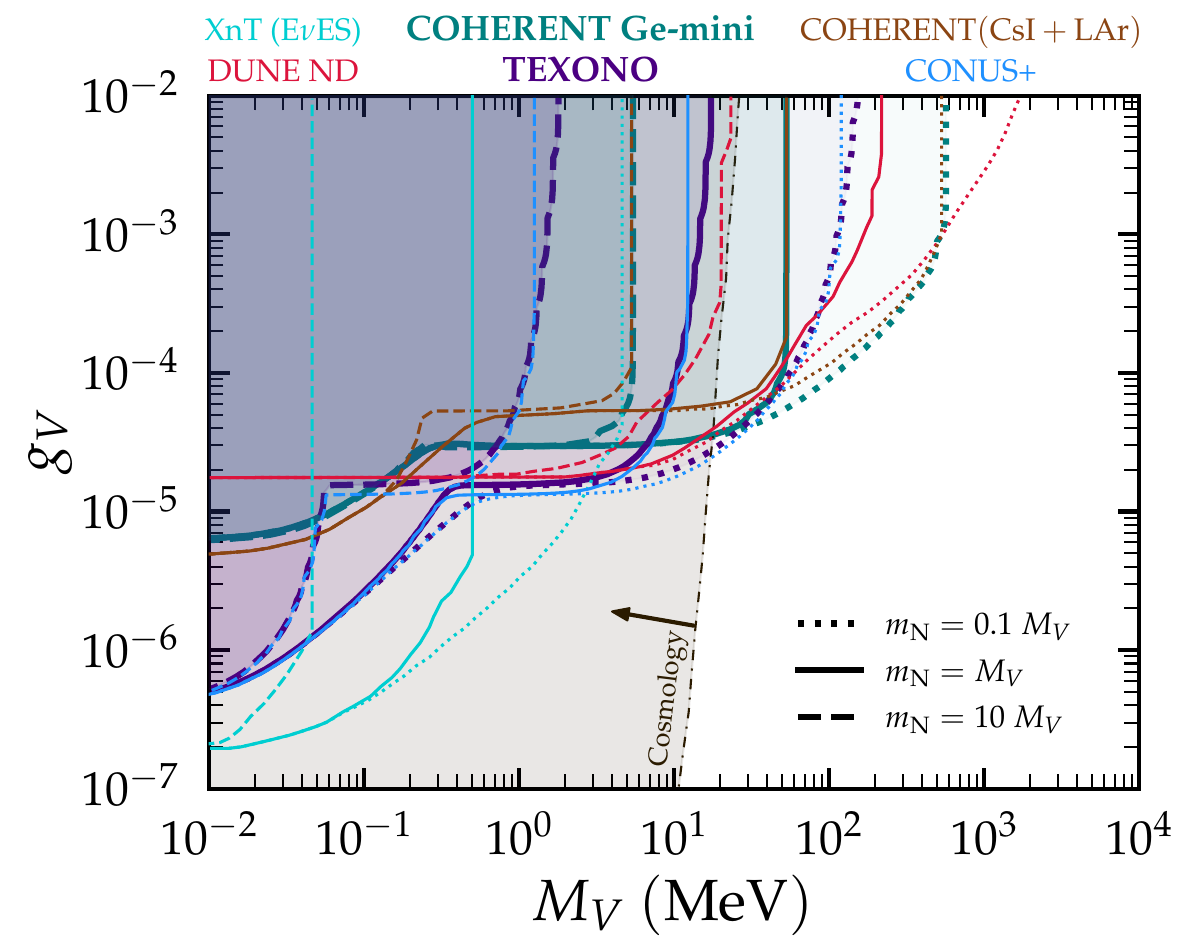}
        \includegraphics[width=0.49\textwidth]{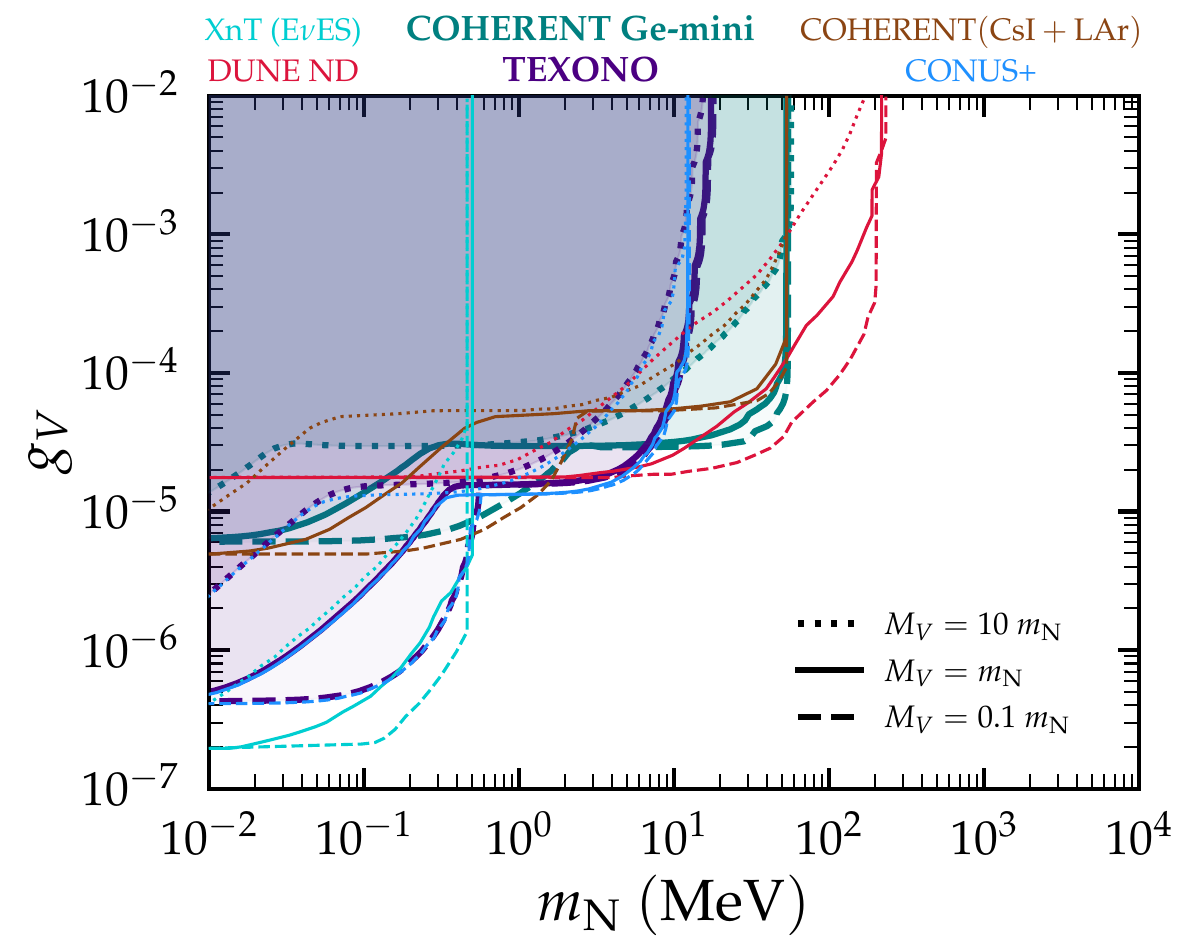}

    \caption{\footnotesize $90\%$ CL exclusion limits on SNL upscattering via generalized interactions for different benchmark choices relating the SNL and mediator masses. The upper and lower panels correspond to the scalar and vector mediator scenarios, respectively. In the left panel of each row, the limits are shown in the $(M_X,\textsl{g}_X)$ plane for the benchmark choices $m_{\rm N}=0.1M_X$ (dotted), $m_{\rm N}=M_X$ (solid), and $m_{\rm N}=10M_X$ (dashed). In the corresponding right panels, the limits are presented in the $(m_{\rm N},\textsl{g}_X)$ plane for the benchmark choices $M_X=10m_{\rm N}$ (dotted), $M_X=m_{\rm N}$ (solid), and $M_X=0.1m_{\rm N}$ (dashed).}
    \label{Fig:Upscattering_Appendix}
\end{figure}

In Sec.~\ref{Sec:Results}, the sensitivities to SNL upscattering mediated by scalar and vector interactions are presented for the representative benchmark $m_{\rm N}=M_X$. To illustrate the dependence of the limits on the assumed relation between the mediator and SNL masses, Figure~\ref{Fig:Upscattering_Appendix} shows the corresponding exclusion limits for the benchmark choices $m_{\rm N}=\{0.1,\,1,\,10\}\times M_X$. For each interaction, the left panels display the limits in the $(M_X,\textsl{g}_X)$ plane, while the right panels present the same results in the complementary $(m_{\rm N},\textsl{g}_X)$ parameter space. In the $(M_X,\textsl{g}_X)$ representation, the dotted, solid, and dashed curves correspond to $m_{\rm N}=0.1M_X$, $m_{\rm N}=M_X$, and $m_{\rm N}=10M_X$, respectively. Conversely, in the $(m_{\rm N},\textsl{g}_X)$ representation, the dotted, solid, and dashed curves correspond to $M_X=10m_{\rm N}$, $M_X=m_{\rm N}$, and $M_X=0.1m_{\rm N}$, respectively. In the $(m_{\rm N},\textsl{g}_X)$ parameter space, the exclusion contours terminate at the kinematic upper limit on the SNL mass for each experiment, which is independent of the adopted benchmark relation between $m_{\rm N}$ and $M_X$. Consequently, in the $(M_X,\textsl{g}_X)$ representation, the corresponding cutoff is translated to the mediator mass axis according to the adopted benchmark relationship between $m_{\rm N}$ and $M_X$.

\FloatBarrier

\bibliographystyle{utphys}
\bibliography{bibliography}

\end{document}